\documentclass[fleqn,usenatbib]{mnras}

\usepackage[T1]{fontenc}

\DeclareRobustCommand{\VAN}[3]{#2}
\let\VANthebibliography\thebibliography
\def\thebibliography{\DeclareRobustCommand{\VAN}[3]{##3}\VANthebibliography}

\usepackage{multirow}
\usepackage[font=small,skip=-5pt]{caption}
\usepackage{soul} 
\usepackage{xargs} 
\newcommandx{\mia}[2][1=]{\todo[linecolor=yellow,backgroundcolor=yellow!25,bordercolor=yellow,inline,#1]{#2}}
\newcommandx{\djbs}[2][1=]{\todo[linecolor=yellow,backgroundcolor=red!25,bordercolor=red,inline,#1]{#2}}

\usepackage{graphicx}	
\usepackage{amsmath}	
\newcommand{\angstrom}{\text{\normalfont\AA}}
\usepackage{amssymb}	
\usepackage{newtxtext,newtxmath}

\title[The radio-loudness of SDSS quasars]{Exploring the radio-loudness of SDSS quasars with spectral stacking}

\author[M. I. Arnaudova et al.]{M. I. Arnaudova,$^{1}$\thanks{E-mail: m.arnaudova@herts.ac.uk}
D. J. B. Smith$^{1}$,
M. J. Hardcastle$^{1}$, 
S. Das$^{1}$,
A. Drake$^{1}$,
K. Duncan$^{2}$,
G. G$\ddot{\rm{u}}$rkan $^{1, 3, 4}$, 
\and M. Magliocchetti$^{5}$, 
L. K. Morabito$^{6, 7}$, 
J. W. Petley$^{6}$, 
S. Shenoy$^{1}$ 
and C. Tasse$^{8, 9}$\\
$^{1}$Centre for Astrophysics Research, University of Hertfordshire, College Lane, Hatfield AL10 9AB, UK\\
$^{2}$Institute for Astronomy, Royal Observatory, Blackford Hill, Edinburgh, EH9 3HJ, UK
\\
$^{3}$Astronomy, ATNF, PO Box 1130, Bentley, WA 6102, Australia\\ 
$^{4}$Thüringer Landessternwarte, Sternwarte 5, 07778 Tautenburg, Germany\\
$^{5}$INAF - IAPS, Via Fosso del Cavaliere 100, 00133, Rome, Italy \\
$^{6}$Centre for Extragalactic Astronomy, Department of Physics, Durham University, Durham DH1 3LE, UK \\
$^{7}$Institute for Computational Cosmology, Department of Physics, University of Durham, South Road, Durham DH1 3LE, UK\\
$^{8}$GEPI $\&$ ORN, Observatoire de Paris, Université PSL, CNRS, 5 Place Jules Janssen, 92190 Meudon, France\\
$^{9}$Department of Physics $\&$ Electronics, Rhodes  University, PO Box 94, Grahamstown, 6140, South Africa \\
}
\date{Accepted XXX. Received YYY; in original form ZZZ}

\pubyear{2024}

\begin{document}

\label{firstpage}
\pagerange{\pageref{firstpage}--\pageref{lastpage}}
\maketitle

\begin{abstract}
We use new 144\,MHz observations over 5634\,deg$^2$ from the LOFAR Two-metre Sky Survey (LoTSS) to compile the largest sample of uniformly-selected, spectroscopically-confirmed quasars from the 14th data release of the Sloan Digital Sky Survey (SDSS-DR14). Using the classical definition of radio-loudness, $R=\log(L_{\rm{1.4GHz}}/L_{i})$, we identify 3,697 radio-loud (RL) and 111,132 radio-quiet (RQ) sources at $0.6<z<3.4$. To study their properties, we develop a new rest-frame spectral stacking algorithm, designed with forthcoming massively-multiplexed spectroscopic surveys in mind, and use it to create high signal-to-noise composite spectra of each class, matched in redshift and absolute $i$-band magnitude. We show that RL quasars have redder continuum and enhanced [O\,\textsc{ii}] emission than their RQ counterparts. These results persist when additionally matching in black hole mass, suggesting that this parameter is not the defining factor in making a QSO radio-loud. We find that these features are not gradually varying as a function of radio-loudness but are maintained even when probing deeper into the RQ population, indicating that a clear-cut division in radio-loudness is not apparent. Upon examining the star formation rates (SFRs) inferred from the [O\,\textsc{ii}] emission line, with the contribution from AGN removed using the [Ne\,\textsc{v}] line, we find that RL quasars have a significant excess of star-formation relative to RQ quasars out to $z=1.9$ at least. Given our findings, we suggest that radio-loud sources either preferably reside in gas-rich systems with rapidly-spinning black holes, or represent an earlier obscured phase of QSO evolution.
\end{abstract}




\section{Introduction}

The most luminous manifestations of active galactic nuclei (AGN) are quasi-stellar objects (QSOs), also known as quasars, whose bolometric luminosity can reach up to 10$^{47-48}$ erg s$^{-1}$ (e.g. \citealt{rakshit2020spectral}; \citealt{shen2020bolometric}). About 5-10$\%$ of these optically-selected sources are found to emit strongly in the radio band, likely due to the presence of relativistic jets (\citealt{urry1995unified}), while the remaining 90$\%$ are only weak radio sources, whose emission could be purely a result of star formation (e.g. \citealt{kimball2011two}; \citealt{condon2013active}; \citealt{kellermann2016radio}). This division into radio-loud (RL) and radio-quiet (RQ) quasars raises the question of whether these two types of objects represent physically distinct populations or different evolutionary stages of a single one. 

To provide an answer to this question, studies have looked into the distribution of the radio-loudness parameter ($R$; the ratio of radio to optical flux density or luminosity).
Some claim that this distribution is bimodal (e.g. \citealt{kellermann1989vla}; \citealt{ivezic2002optical}; \citealt{white2007signals}), whilst others present evidence against this bimodality (e.g. \citealt{cirasuolo2003radio,cirasuolo2003there}; \citealt{balokovic2012disclosing}; \citealt{gurkan2019lotss}; \citealt{macfarlane2021radio}). Part of the reason for these contradictory results could be due to the definition of $R$.  
For example, the use of optical and radio information leads to inhomogeneous samples as a result of different selection effects. In addition, the $R$ parameter is calculated using different bands, depending on data availability, which may not give consistent results \citep{ivezic2002optical}. Finally, both the optical and radio emission could be contaminated by the host galaxy, while the radio emission could be further complicated by the jet power's dependence on the environment, time and Doppler boosting (e.g. \citealt{liu2006jet}; \citealt{gurkan2019lotss}; \citealt{radcliffe2021radio}).

Another debated issue involves the source of radio emission in RQ quasars. In star-forming galaxies, the radio emission is associated with star formation through free-free emission from H${\rm{II}}$ regions and synchrotron radiation from electrons accelerated to relativistic speeds by supernova remnants \citep{condon1992radio}. This leads to the question of whether star formation in the host galaxy is sufficient to account for the observed radio emission from RQ quasars. Some studies find that SF is enough (e.g. \citealt{kimball2011two}; \citealt{condon2013active}), whilst others argue that it must come from AGN, in the form of small-scale jets, AGN-driven winds or disc coronal activity (e.g. \citealt{laor2008origin}; \citealt{zakamska2016star}; \citealt{white2015radio}; \citealt{symeonidis2016agn}; \citealt{white2017evidence}; \citealt{morabito2022identifying}). 

The nature of jet production in RL quasars is also not clear. Following the work of \cite{blandford1977electromagnetic}, some propose that the black hole (BH) spin plays a vital role in powering radio jets (e.g. \citealt{wilson1994difference}; \citealt{sikora2007radio}; \citealt{tchekhovskoy2010black}). However, due to the extreme difficulty in measuring this quantity, it is challenging to test this model observationally. Another potential physical parameter involved in determining the jet power and thus the distinction between RL/RQ quasars is the BH mass. While some authors find that radio-loudness strongly depends on BH mass (e.g. \citealt{gu2001masses}; \citealt{dunlop2003quasars}; \citealt{mclure2004relationship}; \citealt{best2005host}; \citealt{metcalf2006role}; \citealt{chiaberge2011origin}), others have found only a weak dependence or no dependence at all (e.g. \citealt{ho2002relationship}; \citealt{shankar2010relative}; \citealt{gurkan2019lotss}; \citealt{macfarlane2021radio}). 

To investigate these problems, \cite{gurkan2019lotss} combined a sample of optically-selected quasars from the fourteenth data release of the Sloan Digital Sky Survey (SDSS; \citealt{paris2018sloan}) with radio observations from the first data release of the LOFAR Two-metre Sky Survey (LoTSS; \citealt{shimwell2017lofar}) over the HETDEX spring field \citep{hill2008hobby} and the LOFAR H-ATLAS/NGP survey \citep{hardcastle2016lofar}. With the high sensitivity and wide areal coverage of LoTSS, the authors were able to study the dependence of the radio-loudness parameter on galaxy properties such as redshift, bolometric luminosity, radio luminosity, BH mass and Eddington ratio. They found that quasars exhibit a wide continuum of radio properties, with no clear signatures of a bimodality. Given these results, the authors favoured the scenario where both AGN jets and SF contribute to the radio emission in quasars such that there is no RL/RQ dichotomy, but rather a smooth transition between the dominance of these two processes.

Recently, \cite{macfarlane2021radio} built upon these results by developing a numerical model of the radio flux densities of quasars, in which the radio emission of every quasar consists of two components: AGNs (jets) and SF. This model, coupled with Monte Carlo simulations, allowed the authors to create quasar mock samples and compare them with observations. Their results were found to be in excellent agreement with the observed radio flux distributions of $\sim$42,000 SDSS quasars as measured in LoTSS DR1 across several redshift and absolute $i$-band magnitude ranges. This is consistent with a model in which jet production is present in all quasars with a different powering efficiency such that it leads to a smooth transition between the RQ and RL quasar regimes.

Our work takes a different approach by developing a spectral stacking algorithm and using it with the much larger LoTSS DR2 sample. With its extensive coverage of 5634 square degrees, LoTSS DR2 provides a much larger observational volume, resulting in a substantial increase of RL quasars. Employing our stacking techniques on this expanded dataset allows us to systematically create composite spectra for each radio class in a given parameter regime (i.e. redshift, $i$-band luminosity and black hole mass). This approach enables us to thoroughly explore the continuum and emission line properties of quasars, while also determining the influence of key physical parameters such as black hole mass and Eddington ratio. 
Therefore, although radio-loudness may not correspond to a physical property of QSOs, it can be useful for identifying sources with powerful jets, and high S/N ratio spectra provide an excellent way to investigate their properties.
This stacking algorithm is designed with the upcoming WEAVE-LOFAR survey \citep{smith2016weave} in mind, which will provide over a million spectra of LoTSS targets selected at 144 MHz. Such a tool is necessary as a result of the survey's radio selection criteria, which produces samples dominated by AGN and/or ongoing star formation. The spectra of such sources are rich in emission lines which allow us to robustly determine the redshifts, but a continuum detection is not always available (e.g. for faint star-forming galaxies). Stacking such sources together, however, allows us to statistically detect the continuum and thus recover spectral features otherwise indistinguishable in individual detections. Furthermore, stacking sources of different demographics will enable us to create a large library of high resolution templates that will help improve the WEAVE-LOFAR survey's redshift estimates.

The structure of this paper is as follows. Section \ref{sec:data} provides details of the spectroscopic and radio data used in this study, along with the methodology for the sample selection and matching process. In section \ref{sec:methods}, we describe the spectral stacking technique employed for comparing the radio classes of QSOs. Subsequently, in sections \ref{sec:results} and \ref{sec:spec_prop}, we create high signal-to-noise (S/N) composite spectra of QSOs and investigate potential factors contributing to the observed effects between them. Finally, section \ref{sec:discussion} discusses possible explanations and section \ref{sec:conclusion} summaries our main results.
Throughout this work, we use vacuum wavelengths and a flat $\Lambda$CDM cosmology with
$\Omega_{\Lambda}$ = 0.7, $\Omega_{M}$ = 0.3 and $H_{0}$ = 70 km s$^{-1}$ Mpc$^{-1}$.

\section{Data}\label{sec:data}

\subsection{Sloan Digital Sky Survey}\label{sec:SDSS}

The spectroscopic data used in this work are taken from the fourteenth data release of the Sloan Digital Sky Survey Quasar Catalogue (SDSS-DR14Q), which is fully described by \citet{paris2018sloan}. This catalogue includes all spectroscopically confirmed quasars from SDSS-I/II, SDSS-III/BOSS and SDSS-IV/eBOSS programmes, resulting in a sample of 526,356 objects over a region of 9376 $\deg^{2}$ as shown in Figure \ref{fig:sky_map}. 

SDSS-I/II contains 79,847 quasars with $i$-band absolute magnitudes brighter than $M_{i} [z=2] = -22.0$ over a wide redshift range of $0.065 < z < 5.46$. The targetting algorithm used to obtain these sources has been updated throughout the years and more information about it is provided by \citet{richards2002spectroscopic} and \citet{schneider2010sloan}. The spectra are produced by a pair of multi-object spectrographs (SDSS) that have a total of 640 fibres with an entrance diameter of 3$\arcsec$ and an average resolving power of $\Delta \lambda/\lambda \approx 2000$ across the vacuum wavelength range from 3800 $\angstrom$ to 9200 $\angstrom$.

The spectra for SDSS-III/BOSS and SDSS-IV/eBOSS, on the other hand, are produced by the upgraded pair of spectrographs used for the Baryon Oscillation Spectroscopic Survey (BOSS). These spectrographs benefit from having a total of 1000 fibres with a smaller entrance diameter of 2$\arcsec$, an extended vacuum wavelength coverage of $3600 - 10,400$ $\angstrom$, and a resolving power higher in the red channel and lower in the blue channel compared to the SDSS \citep{smee2013multi}.
These two programs use multiple target selection algorithms to detect much fainter quasars ($M_{i} [z=2]= -20.5$) at redshift ranges of $2.15 < z < 3.5$ and $0.9 < z < 2.2$, respectively (for further details see \citealt{ross2012sdss} and \citealt{myers2015sdss}). 
As a result of these differences and the considerable size of 446,781 for the latter two programs, we use only the BOSS spectra in our analysis.

Complementary to this SDSS-DR14Q catalogue is the value-added catalogue by \citet{rakshit2020spectral} which includes continuum and line property measurements, including bolometric luminosity ($L_{\mathrm{bol}}$), derived virial black hole mass ($M_{\mathrm{BH}}$) and Eddington ratio ($R_{\mathrm{edd}}$) estimates. This was done using the publicly available spectral fitting code \textsc{\texttt{PyQSOFit}} \citep{guo2018pyqsofit}, which uses two independent sets of eigenspectra -- pure galaxy \citep{yip2004distributions} and pure quasar \citep{yip2004spectral} -- to decompose the spectrum into host galaxy and quasar contribution.
This decomposition is particularly important in low-redshift quasars ($z<0.8$), where the stellar contribution can be significant (e.g. \citealt{yue2018sloan}; \citealt{shen2019sloan}; \citealt{rakshit2020spectral}). Briefly, the continuum of the host free spectrum is modelled by a combination of power-law, Fe\,{\sc ii} and Balmer continuum models, whereas the emission lines are fitted with Gaussian distributions. The bolometric luminosity is computed from $L_{5100}$ ($z$ < 0.7), $L_{3000}$ (0.7 $\leq z$ < 1.9) and $L_{1350}$ ($z \geq$ 1.9) using the bolometric corrections from \cite{richards2006spectral}. 
For the virial black hole mass multiple measurements are included in the catalogue depending on the availability of strong emission lines and various calibrations. In this work, we choose to use the ``fiducial'' estimate calculated based on the  H\,{\sc $\beta$} line ($z<0.8$) and the C\,{\sc iv} line ($z\geq 1.9$) using the calibrations from \cite{vestergaard2006determining}, and the Mg\,{\sc ii} line ($0.8\leq z < 1.9$) using the calibration from \cite{vestergaard2009mass}. These measurements are also used to calculate the Eddington luminosity and thus the Eddington ratio, which is used as a proxy for the accretion rate.

\subsection{LOFAR Two-metre Sky Survey}\label{sec:LoTSS}

\begin{figure}
	\includegraphics[width=\columnwidth]{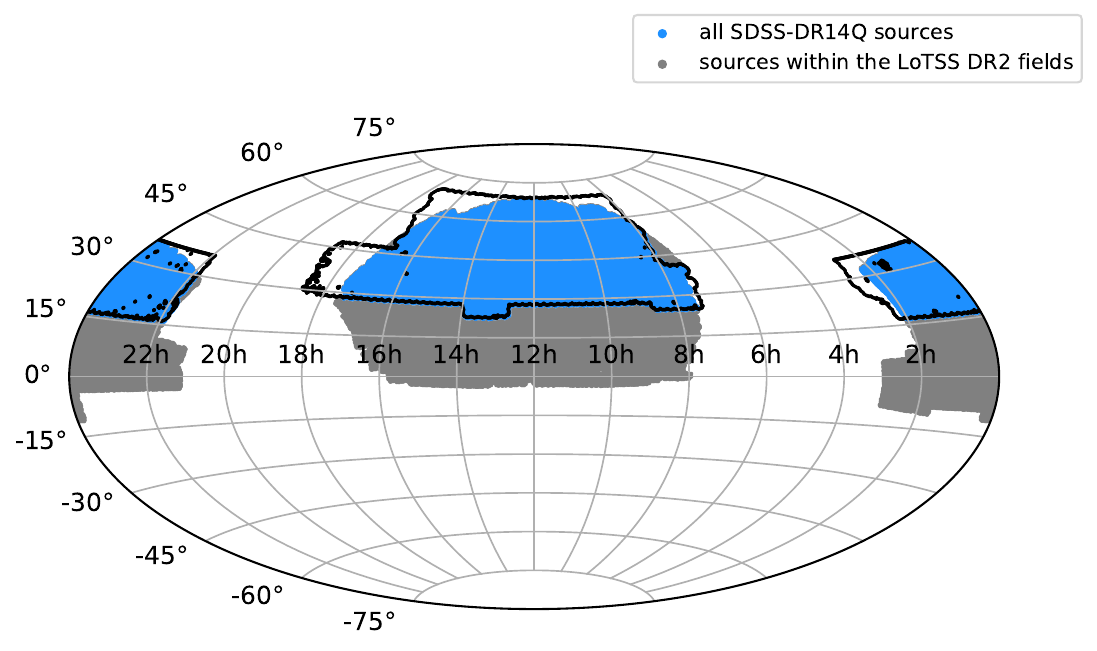}
    \caption{Sky coverage of SDSS-DR14Q and LoTSS DR2. The grey region indicates all the sources from the SDSS-DR14Q catalogue, whereas the light blue region - those located in the LoTSS DR2 footprint, where cross-matching has been performed.}
    \label{fig:sky_map}
\end{figure}

The radio data used in this work are taken from the second data release of the LOFAR Two-metre Sky Survey (LoTSS DR2; \citealt{shimwell2022lofar}). This data release covers 5634 deg$^2$ of the northern sky (see footprint in Figure \ref{fig:sky_map}) at a resolution of 6$\arcsec$ and a median rms sensitivity of 83 $\mu$Jy beam$^{-1}$, providing over 4 million radio sources, the vast majority of which have never been detected at radio wavelengths before. In addition to the large area and high sensitivity, LoTSS benefits from using the low radio central frequency of 144 MHz which amongst other advantages reduces the effects from Doppler boosting in jetted sources.

In preparation for the upcoming WEAVE-LOFAR survey, \cite{hardcastle2023lofar} have created a preliminary cross-matched catalogue containing 296,921 SDSS counterparts by using a combination of statistical methods and visual classification in a similar manner to \cite{kondapally2021lofar}.
To summarise, for smaller, more compact radio sources, the likelihood ratio method is applied (e.g. \citealt{richter1975search}; \citealt{de1977westerbork}; \citealt{sutherland1992likelihood}). This statistical approach uses both colour and magnitude information as described by \cite{williams2019lofar} in order to maximise the number of identifications, as well as increase the robustness of the cross-matching. For larger sources with extended radio emission, the web-based interface for visual inspection - the Radio Galaxy Zoo: LOFAR, is needed. Here, the user can perform identification and association for a given radio source by making use of available multi-wavelength images. Finally, for everything in between a decision tree is used to combine the two methods. 

\subsection{Final Sample}\label{sec:Final Sample}

Starting with a sample of 446,781 BOSS quasars from the DR14Q catalogue, we identify those located in the LoTSS DR2 area with the use of a Multi-Order Coverage (MOC) map as seen in Figure \ref{fig:sky_map}. This results in 265,578 objects, out of which 33,968 are part of the LoTSS DR2 optical cross-matched catalogue. 
Next, we remove spectra with bad plate quality and a non-zero quality flag for $M_{\mathrm{BH}}$ and $L_{\mathrm{bol}}$ to ensure reliability of results. Finally, to mitigate any selection bias, we use the target selection flags to discard objects targeted by SDSS solely based on their radio or X-ray emission, as well as any found to have time variability. 
The final sample consists of 189,680 quasars, out of which 123,742 are part of the BOSS/eBOSS homogeneously-selected CORE sample. The CORE sample is created by using a single target selection algorithm, the Extreme Deconvolution (XDQSO; \citealt{bovy2011think}), which is designed to meet science goals (such as clustering and LF measurements) which require a uniform selection, and so is ideal for our purposes.

To separate quasars into radio-loud and radio-quiet, we adopt the standard definition of the radio-loudness parameter ($R$): the logarithm of the ratio of 1.4 GHz radio luminosity ($L_{1.4\mathrm{GHz}}$) to optical $i$-band luminosity ($L_{i}$), where $R = 1$ marks the boundary (e.g. \citealt{balokovic2012disclosing}). 
To calculate the $k$-corrected $L_{1.4\mathrm{GHz}}$ for radio-detected QSOs, we use the integrated 144 MHz flux density ($S_{144\mathrm{MHz}}$) from the LoTSS DR2 catalogue, a radio spectral index of $\alpha_{\mathrm{rad}} = -0.7$ and the sources' spectroscopic redshifts reported in the DR14Q catalogue. The same calculation is applied to the radio-undetected QSOs in order to obtain a $5\sigma$ upper limit for $L_{1.4\mathrm{GHz}}$, where $S_{144\mathrm{MHz}}$ is taken to be $5\times$ the local rms value taken directly from the LoTSS rms maps at the coordinates of the quasars given by SDSS DR14Q.

For the optical luminosity, we take the absolute $i$-band magnitude from the DR14Q catalogue which is $k$-corrected to $z=2$ and use the following conversion given by \cite{richards2006spectral} to obtain a $k$-corrected estimate to $z=0$:
\begin{equation}
M_{i}[z=0] = M_{i}[z = 2] + 2.5(1+\alpha_{\mathrm{opt}})\log(1+z).
\end{equation}
Here $z=2$ and we adopt an optical spectral index of $\alpha_{\mathrm{opt}}$ = -0.5. To obtain the $i$-band luminosity, we use the simple relationship between luminosity and magnitude, where the solar luminosity ($3.827\times10^{26}$ W) and solar $i$-band absolute magnitude (4.58) were used. This gives us a total of 3,697 RL and 111,132 RQ quasars. The remaining sources are discarded as their $5\sigma$ upper limit puts them in the radio-loud regime, such that we are unable to classify them with confidence. 

\begin{figure}
	\includegraphics[width=\columnwidth]{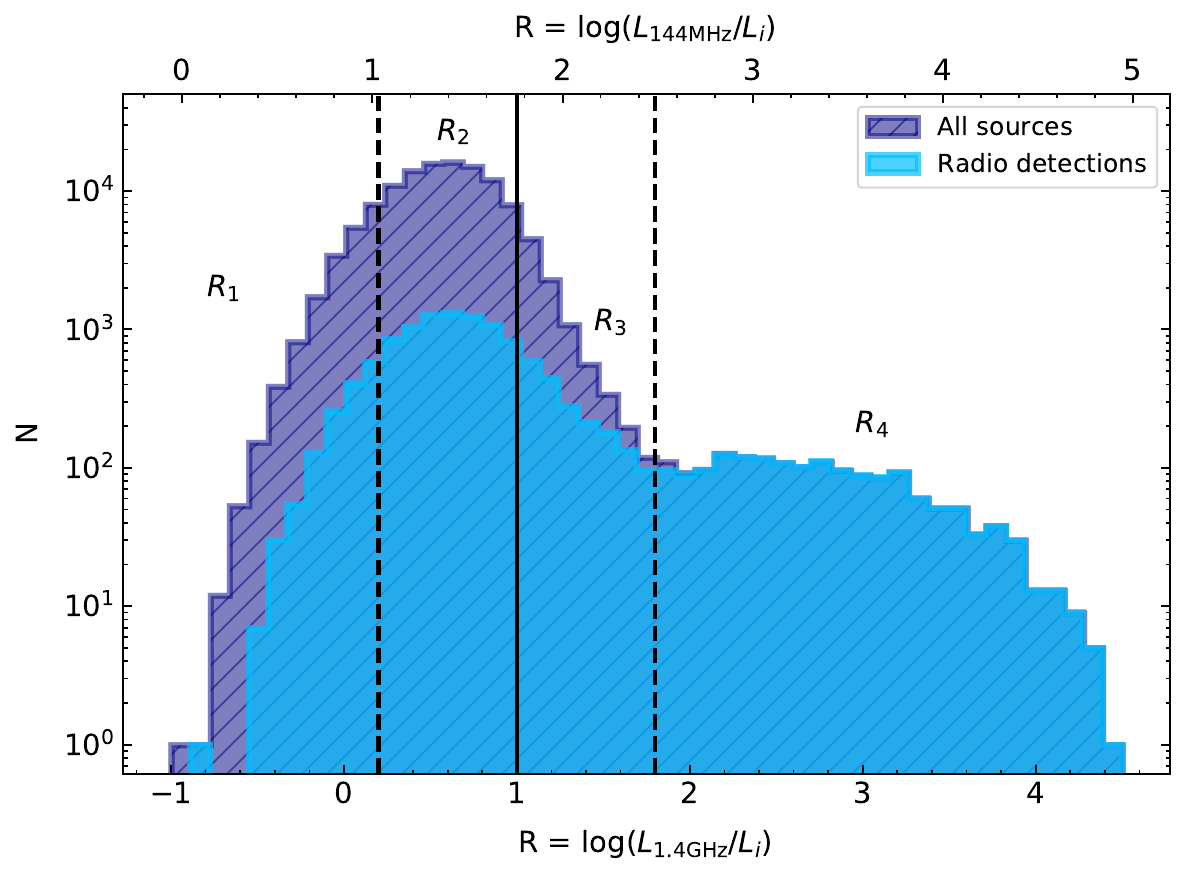}
    \caption{The radio-loudness parameter for all sources (dark blue) and only radio-detections (light blue). The solid black line denotes the standard division line for radio-loud and radio-quiet QSOs, whereas the dashed black lines are additional divisions we make to define classes as a function of $R$ as denoted.}
    \label{fig:R}
\end{figure}

The radio-loudness distribution of all sources, including a subset of radio detections, is presented in Figure \ref{fig:R} (where we have also included a second horizontal axis to show $R$ in terms of the $L_{\rm{144MHz}}$ for comparison). It is evident that in both cases the conventional demarcation line ($R=1$) does not represent any particularly significant level for the sample, however we have repeated the analysis considering different values (as denoted in Figure \ref{fig:R}) finding that our results are qualitatively unchanged. Doing these tests gives additional advantages of further testing whether there is a gradual change between the two populations as found in previous studies (e.g. \citealt{gurkan2019lotss}; \citealt{macfarlane2021radio}), as well as avoiding  biases that may result from assuming a single radio spectral index.

\subsection{The Matching Process}\label{sec:matching}

To make a robust comparison between quasar populations, we develop a method to create samples matched in 2D and 3D parameter space. For clarity, in what follows we describe the method for the RL and RQ population, where we create samples matched in $z$, $M_{i}$ and $z$, $M_{i}$, $M_{\rm{BH}}$ to use in sections \ref{sec:comparison} and \ref{sec:MBH}. However, it is important to note that this method can be adapted to various other scenarios, as demonstrated throughout this study.

The 2D matching process involves generating a 2D histogram of the RL QSOs, and identifying the number of RQ that fall within the same $z$ and $M_{i}$ bin. For the one-to-one case, this entails choosing the same number of RQ counts as found for the RL 2D-histogram, such that if one bin is populated by 100 RL sources then we randomly select only 100 RQ counterparts. However, as our RQ sample is considerably larger, we instead choose to normalise the RL 2D-histogram by the total number of RL sources, and multiply the normalised counts by the maximum number of RQ QSOs for which we can say that they are drawn from the same $z$ and $M_{i}$ distribution as the RL population. This is done by employing multiple Kolmogorov–Smirnov (K-S) tests until we accept the null hypothesis that the two samples are drawn from the same distribution if the $p$-value is greater than a significance level of $\alpha = 0.05$. This results in a sample of 77,532 RQ QSOs matched in $z$ and $M_{i}$ to our RL sample (hereafter the 2D-matched or $z-M_{i}$ sample).

\begin{figure}
	\includegraphics[width=\columnwidth]{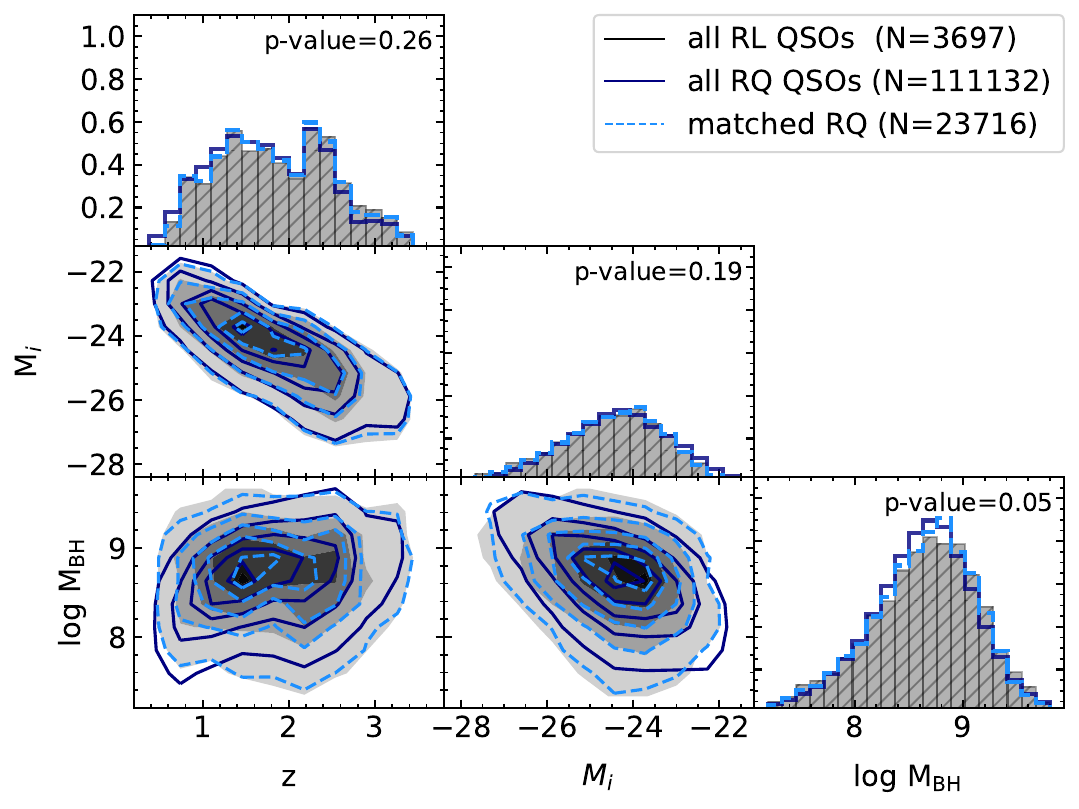}
    \caption{A corner plot summary of the redshift, absolute $i$-band magnitude and black hole mass distributions for the RL and RQ QSOs. The density contours represent the 10, 30, 50, 70 and 90 per cent of the whole (dark blue) and 3D-matched (dashed, light blue) RQ sample, whereas the density map represents the corresponding percentages for the RL sample. The histograms share a common y-axis and indicate the individual $z$, $M_{i}$ and $M_{\rm{BH}}$ distribution, where the quoted p-values in the upper-right corner are obtained by performing a K-S test on the RL and 3D-matched RQ sample.}
    \label{fig:corner_plot}
\end{figure}

Similarly, the 3D matching process to create a sample matched in $z$, $M_{i}$ and $M_{\rm{BH}}$ (hereafter the 3D-matched or $z-M_{i}-M_{\rm{BH}}$ sample) involves creating a normalised 3D-histogram of the RL population, and again using multiple K-S tests to determine the maximum number of RQ sources. The effectiveness of this matching procedure can be seen in Figure \ref{fig:corner_plot}, where we compare the individual $z$, $M_{i}$ and $M_{\mathrm{BH}}$ distributions, along with contours of the 2D-distributions of our 3D-matched sample of 23,716 RQ QSOs with the whole RL and RQ samples.

\section{The Spectral Stacking Method}\label{sec:methods}

\begin{figure*}
\centering
\includegraphics[width=.9\textwidth]{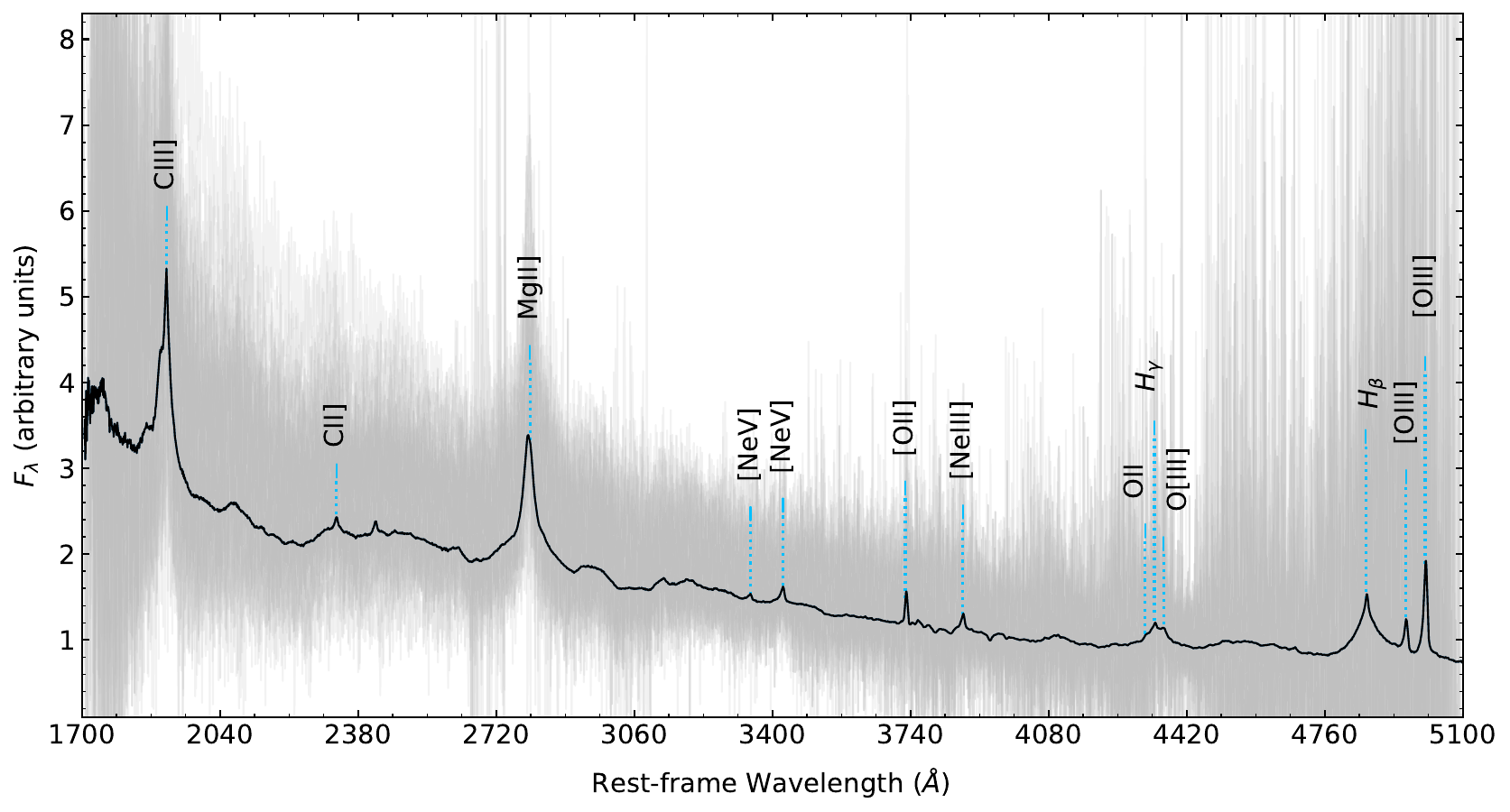}
    \caption{An example composite spectrum of RQ QSOs. The average spectrum is indicated by the thick black line, while the grey region indicates the individual normalised, resampled spectra that went in to the stack. The labelled dotted blue lines indicate prominent emission lines visible in the stacked spectrum.}
    \label{fig:stack}
\end{figure*}

\subsection{The Stacking Technique}\label{sec:stacking}

To create high S/N composite spectra of quasars we use the following method. First, we correct for Galactic extinction in the observed-frame by using the re-calibrated reddening data, $E(B-V)$, from \citet*{schlegel1998maps}, along with the Milky Way reddening curve from \cite{fitzpatrick1999correcting} for an extinction-to-reddening ratio of $R_{V}=3.1$.

Next, we shift the spectra to the rest-frame by using the spectroscopic redshifts from the SDSS-DR14Q catalogue, and resample onto a common wavelength grid, predetermined by the minimum and maximum value of the rest-frame wavelengths sampled, along with the choice of sampling input in the algorithm (e.g. 1 $\angstrom/\text{pixel}$). The resampling is performed using the \textsc{\texttt{SpectRes}}:Simple Spectral Resampling tool \citep{carnall2017spectres}, where the old wavelength grid is cross-matched with the new one, such that if an old wavelength bin spreads between multiple new ones, the flux density value associated with it is distributed in the new grid based on the fractional coverage. This ensures that the flux density is conserved.
To normalise the resampled spectra, we divide through each spectrum by its median, computed at the reddest possible end of the area where all spectra populate the common wavelength grid, where prominent emission lines are masked out. Doing this ensures that the overall spectral shape is preserved, in the sense that we are able to recover stellar population synthesis models once the uncertainties have been estimated and corrected for bias (see next section for details). In addition, by using the reddest possible common wavelength range, we minimize the effects of extinction.

Stacking the de-redshifted, resampled and normalised spectra is generally done in the literature either by taking the weighted average, where the weights are given by the inverse variance of the spectra, or by taking the median (e.g. \citealt{rowlands2012herschel}; \citealt{zhu2015near}; \citealt{rigby2018magellan}; \citealt{calabro2021vandels}). The weighted average method is found to produce a higher S/N composite spectrum, however, as explained in \cite{calabro2021vandels}, this is the result of a bias towards lower redshifts and/or individual spectra with higher S/N. On that account, we have chosen to build a composite spectrum by taking the median of all normalised flux density values that fall within a given wavelength bin.
To calculate the uncertainty associated with this stack, we use bootstrapping to randomly resample the spectra at each wavelength bin, creating 1000 realisations with the same size as the original sample. The median method is then performed for each realisation to estimate the flux density distribution at a given bin and thus determine the standard deviation.

Figure \ref{fig:stack} presents the result of implementing this method on a sample of RQ quasars, where the individual spectra are shown in grey, and the high resolution stack in black. This shows how powerful spectral stacking
really is in identifying spectral features otherwise undetected in
individual spectra.

\subsection{The Stacking Corrections}\label{sec:corrections}

To evaluate the performance of the stacking procedure described in the previous section, we create a model galaxy spectrum by making use of the stellar population synthesis code from \cite{bruzual2003stellar}, along with equivalent width measurements of nebular emission lines taken from the value added catalogue produced by the Max-Planck-Institute for Astrophysics-John Hopkins University (MPA-JHU; \citealt{brinchmann2004physical}; \citealt{kauffmann2003stellar} and \citealt{tremonti2004origin}). This galaxy template has a broad wavelength range extending from 90 $\angstrom$ to 160 $\mu$m, a resolving power similar to that of SDSS across the range from 3200 to 9500 $\angstrom$ and contains the main nebular emission lines targeted by both WEAVE-LOFAR and SDSS. These features enable us to simulate mock samples of galaxy spectra based on SDSS characteristics across a wide range of redshifts, providing us with a robust way of testing the spectral stacking algorithm and further demonstrating its flexibility in handling different type of spectra (see Figure \ref{fig:BPT stacks} for an example of stacking spectra as a function of BPT classes).

Following a range of tests including continuum recovery at low brightness and performance related to different sampling of spectra (the main motivations for implementing it for the WEAVE-LOFAR survey), we have identified two implications of the stacking method. First of all, it is not always possible for the algorithm to choose a normalisation range without strong spectral features, which causes an offset at the blue end of the composite spectrum. Secondly, the spectra are combined in the rest-frame, where the normalised spectra no longer have the same spectral resolution as a result of the de-redshifting process. This is not accounted for in the bootstrap method, leaving the uncertainties underestimated. To correct for these effects, we include in our stacking algorithm the following procedure. 

\begin{figure*}
\centering
\includegraphics[width=.9\textwidth]{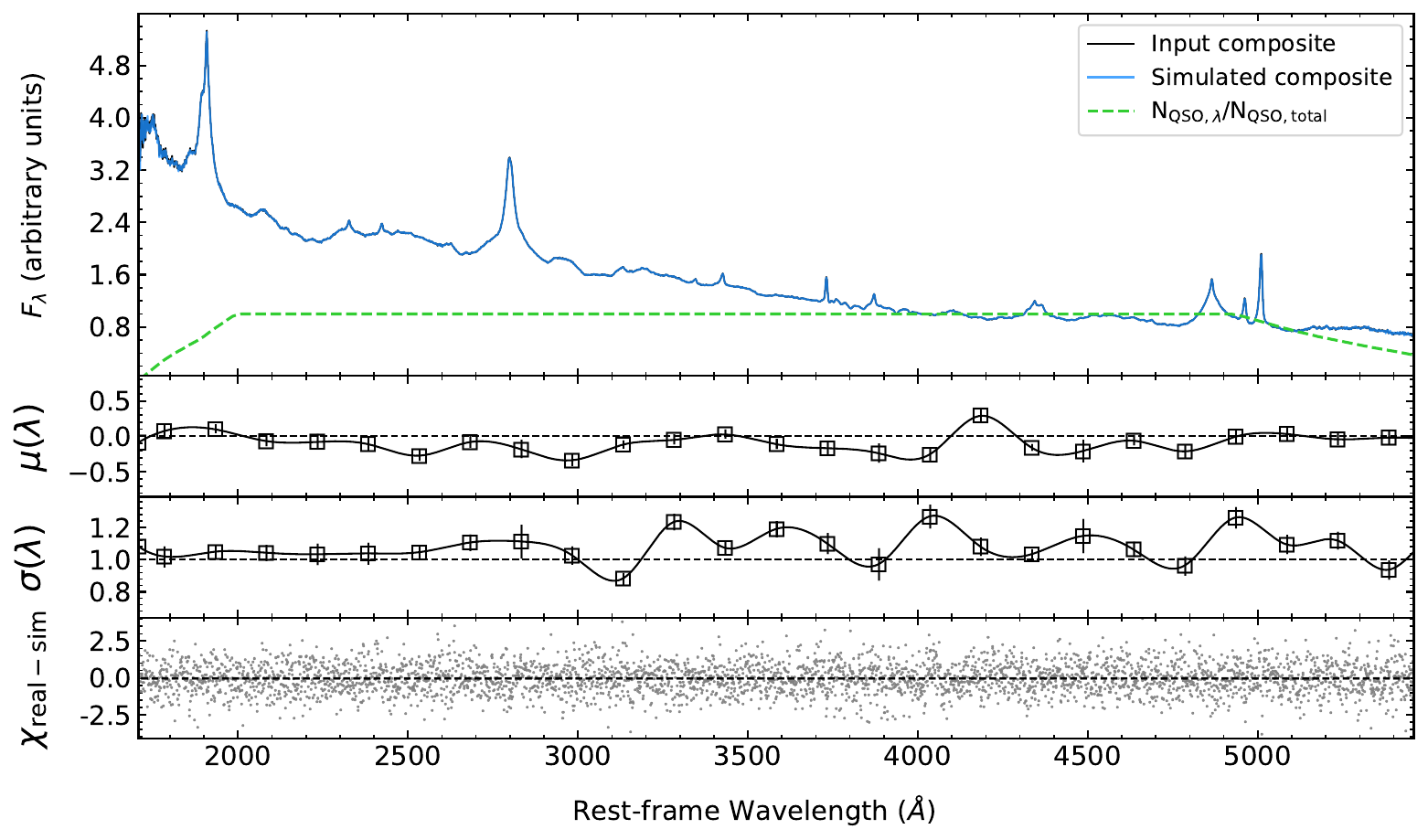}
    \caption{A diagnostic plot of an example composite spectrum of RQ QSOs. The upper panel shows the input (dark blue) and simulated (light blue) composite spectra, along with the number of QSOs that went into them at each rest-frame wavelength divided by the total number of QSOs  (dashed green line). The second and third panel present the best-fitting parameters for the mean and standard deviation for each wavelength slice, while the lower panel presents the corrected residual in units of the propagated uncertainty.}
    \label{fig:correction_plot}
\end{figure*}

For a given spectroscopic sample, we create a composite spectrum (hereafter the input composite spectrum) and use it as a template to simulate quasar spectra based on the characteristics of the original sample such
as redshift, $i$-band magnitude and inverse variance. Next, we stack the simulated spectra to produce a second spectrum (hereafter the simulated composite spectrum) and obtain the residual in uncertainty units, which is defined as:
\begin{equation}
    \chi_{\mathrm{input-sim}}=\frac{F_{\lambda,\mathrm{input}}-F_{\lambda,\mathrm{sim}}}{\sigma_{\lambda,\mathrm{sim}}},
\end{equation}
where $F_{\lambda,\mathrm{input}}$ and $F_{\lambda,\mathrm{sim}}$ are the flux density of the input and simulated composite, respectively and ${\sigma_{\lambda,\mathrm{sim}}}$ is the bootstrap uncertainty for the simulated composite. Provided that the issues discussed above are not present, this quantity would be
normally distributed with a mean of zero and a standard deviation of
one, making it the source of our corrections. To obtain the necessary correction factors, we separate $\chi_{\mathrm{input-sim}}$ into 30 wavelength slices (or less, depending on pixel availability) and for each one we fit a normal distribution. The best-fit parameters are then interpolated with a spline to produce wavelength dependent corrections as indicated in Figure \ref{fig:correction_plot}. 
The best-fit of the mean is used to correct for the offset as a result of the normalisation by subtracting a multiple of it and the uncertainty of the simulated composite from the flux density of the input composite. The best-fit of the standard deviation, on the other hand, is used as a multiplication factor for the uncertainty of the input composite spectrum, such that resolution effects are now taken into account. 

We note that this procedure does not account for different levels of extinction or diversity of samples. However, spectra of QSOs are reasonably homogeneous and the use of large statistical samples binned in a given parameter space is close to homogeneity.

\subsection{Qualitative Template Comparison}

To compare our spectral stacking techniques with the literature, we create RL and RQ composites spanning the broadest possible wavelength coverage. Generating such templates involves using sources at widely different redshifts ($0.5<z<3.5$), which affects our method in two ways. First of all, it has implications for the choice of sources which must be drawn from a similar $z-M_{i}$ space in order to reduce the impact of the Malmquist bias. Secondly, the stacking procedure needs to be performed on smaller redshift ranges as a result of the normalisation process, which requires a common wavelength range free from strong spectral features. To account for this, we select all sources with $-26<M_{i}<-24$ and divide them in redshift bins of size $\Delta z=0.5$ to create six composite spectra per quasar population using the stacking algorithm and the spectral corrections described in sections \ref{sec:stacking} and \ref{sec:corrections}. These stacks are then re-scaled to the composite at the lowest redshift bin and combined into a single template using the inverse-variance weighted average. 

\begin{figure*}
\includegraphics[width=\textwidth]{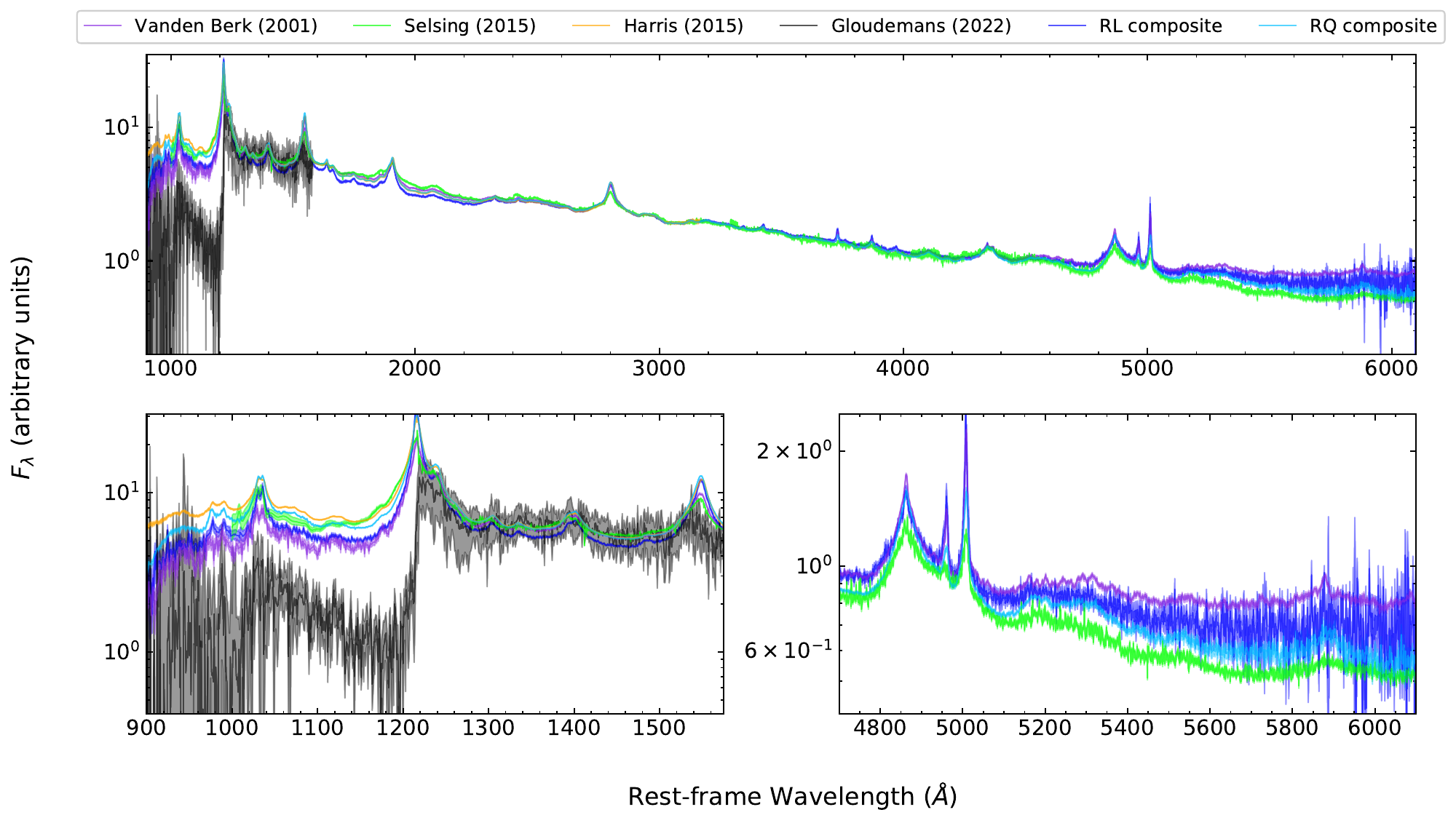}
    \caption{A comparison between various composite spectra as indicated by the legend, where the thick lines indicate the flux density, whereas the shaded area corresponds to $\pm$ the uncertainty. The upper panel presents the full wavelength coverage, where each composite is normalised to the RQ composite at $3000-3600 \angstrom$. The lower left panel presents a zoom-in window indicating the differences bluewards of Ly$\alpha$, whereas the lower right panel - the discrepancy above 5000$\angstrom$ which is likely caused by host galaxy contamination.}
    \label{fig:template}
\end{figure*}

The resulting RL and RQ composites can be seen in Figure \ref{fig:template} where they are compared to the SDSS composite from \cite{berk2001composite}, the X-Shooter composite from \cite{selsing2016x}, the BOSS composite from \cite{harris2016composite} and to a new high-$z$ template that we have constructed by applying our algorithm to the sample of 24 radio-bright quasars (21 out of which are classified as radio-loud) at $4.9< z < 6.6$, discovered by \cite{gloudemans2022discovery}. 
Building such a template presents new challenges, since unlike the homogeneous eBOSS spectra used for the RL\slash RQ templates, the high-$z$ stack requires combining heteroscedastic data obtained from a range of facilities, including the Faint Object Camera and Spectrograph (FOCAS; \citealt{kashikawa2002focas}) on the Subaru Telescope, the LRS2 instrument \citep{chonis2016lrs2} on the Hobby-Eberly Telescope and the Low Resolution Imaging Spectrometer (LRIS; \citealt{oke1995keck}) on the Keck telescope. 
Specifically, and as discussed in section \ref{sec:corrections}, the simulation component of our stacking algorithm requires photometry with a bandpass that overlaps with the observed spectrum, and given that this was unavailable for the \cite{gloudemans2022discovery} sample, we were unable to conduct the simulations and it is therefore likely that the flux density has a normalisation bias with uncertainties systematically underestimated. However, creating and using such a template is beneficial as this is the largest sample of radio-loud quasars at $z>4.9$ and its use demonstrates the flexibility of our stacking algorithm in dealing with non-SDSS spectra.

Despite the difference in selection, we find qualitatively a good agreement in the continuum shape with the various templates. The difference observed redwards of 5000$\angstrom$ amongst the composites is likely caused by various degrees of host galaxy contamination. The SDSS composite from \cite{berk2001composite} includes intrinsically faint sources at low redshifts, making it subject to significant host contamination (e.g. \citealt{glikman2006near}; \citealt{fynbo2012optical}). The selection by \cite{selsing2016x}, on the other hand, is chosen specifically to circumvent this problem, suggesting that both the RL and RQ composite contain low levels of host galaxy contamination. This will be further investigated in section \ref{sec:em}. The disagreement bluewards of Ly$\alpha$, on the other hand, can be explained by the different IGM absorption corrections applied (or the lack thereof). We have not made any corrections, and so the intense drop in flux density for the high-$z$ template is expected considering the redshift range of the sample. The RL composite appears to agree well with the results of \cite{berk2001composite} as seen in the lower left panel of Figure \ref{fig:template}, where no IGM correction is applied as well. Interestingly, the RQ composite appears to be less affected by the Ly$\alpha$-forest absorption as it is found to be in better agreement with the X-shooter and BOSS composite where each spectrum within the composites has been individually corrected.

\begin{figure*}
\includegraphics[width=\textwidth]{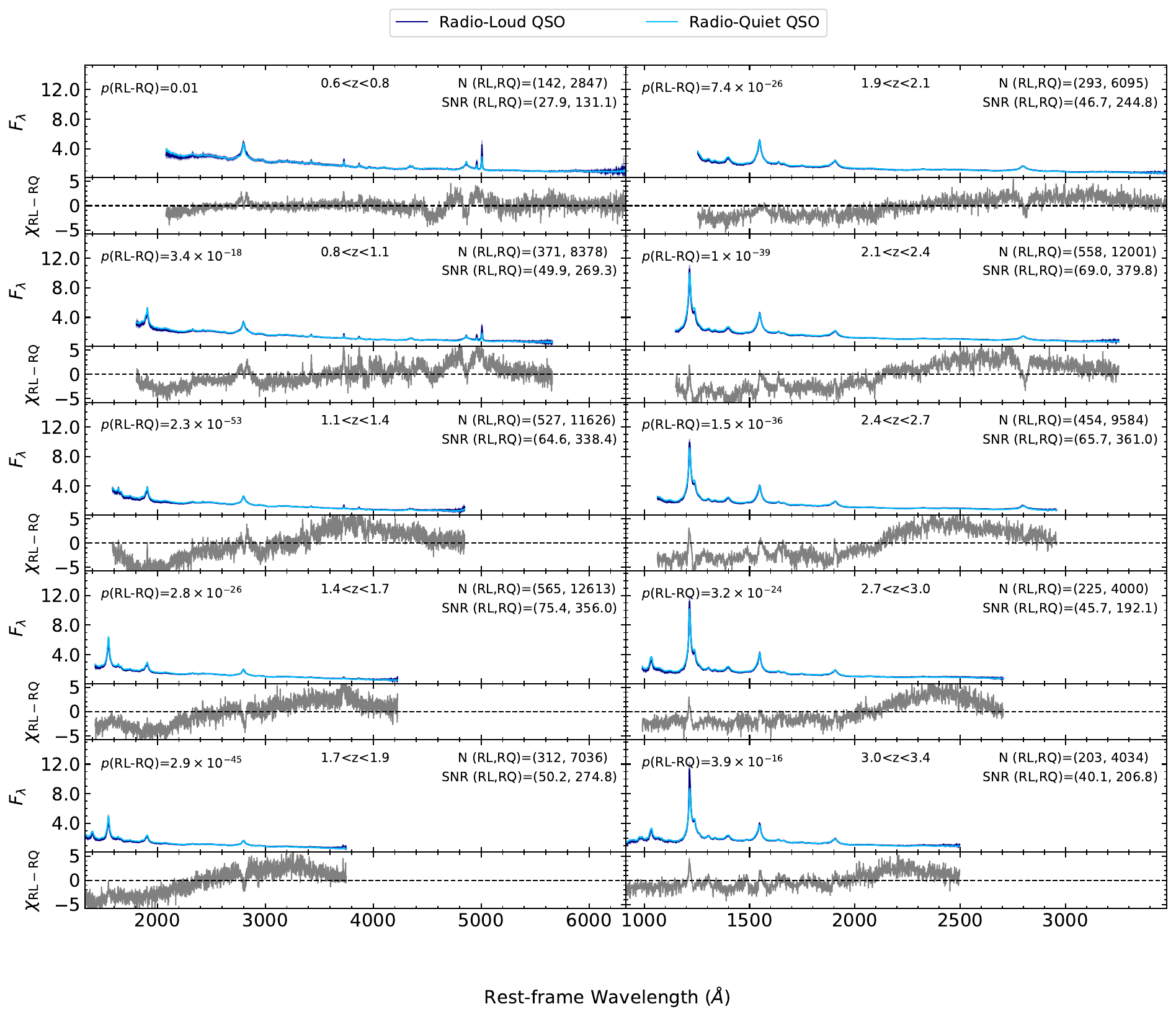}
    \caption{A comparison between the RL and RQ population across ten redshift bins. The upper panel of each bin presents the composite spectra of RL QSOs (dark blue) and RQ QSOs (light blue), whereas the lower panel indicates the residual in units of propagated uncertainty (grey). In each panel, the redshift range is indicated in the upper centre of the plot, while the $p$-values of the null hypothesis test are presented in the legend in the upper left corner. Finally, the number of sources and median S/N is indicated in the legend in the top right corner.}
    \label{fig:RL_vs_RQ}
\end{figure*}

\section{Composite Spectra of Quasars}\label{sec:results}

In this section, we employ our spectral stacking techniques to compare quasars as a function of radio-loudness at different redshift bins. In sections \ref{sec:comparison} and \ref{sec:MBH}, we construct high signal-to-noise (S/N) composite spectra for the classically defined RL and RQ QSOs using the $z-M_{i}$ and $z-M_{i}-M_{\rm{BH}}$ samples to establish whether there are any physical differences between them and investigate the potential influence of the black hole mass. In section \ref{sec:radio bins}, we re-bin our sample into four radio-loudness bins to investigate the presence of a bimodality among the QSOs or a more gradual transition between these populations.

\subsection{High S/N Comparison of RL and RQ QSOs}\label{sec:comparison}

Starting with our $z-M_{i}$ sample, we use the spectral stacking algorithm described in section \ref{sec:stacking} and \ref{sec:corrections} to define high S/N composite spectra across ten redshift bins for each population. The redshift bins are chosen such that they are consistent with the different redshift ranges used to calculate the ``fiducial'' $M_{\rm{BH}}$ from \cite{rakshit2020spectral} and contain comparable numbers of RL sources.

The resulting comparisons for all ten redshift bins are presented in Figure \ref{fig:RL_vs_RQ}. For each bin we present an upper panel including the composite spectra representative of the RL (dark blue) and RQ population (light blue) and a lower panel indicating the residual in units of the propagated uncertainties (grey). We can see that there is a consistent picture emerging across all redshift bins considered - the RL QSOs appear to have on average a redder continuum with notable differences in a number of prominent emission lines.

To quantify the significance of the observed results, we perform the same stacking methods using only the RQ population in order to create Monte Carlo simulations under the null hypothesis that the RL sample is consistent with having been randomly drawn from the RQ sample. For each simulation we take the full sample of RQ sources and randomly draw a sub-sample of size equal to the RL population (hereafter referred to as the RL test sample). The remaining RQ sources are then matched in $z$ and $M_{i}$ as described in section \ref{sec:matching} to create a RQ test sample. Finally, the stacking algorithm is applied on the two test samples to create composite spectra across the same ten redshift bins for which we find a "reduced chi-squared value" of 
\begin{equation}
    \chi^{2}_{\nu} = \frac{\sum_{i=1}^{N} \frac{(F_{RL,i} - F_{RQ,i})^{2}}{(\sigma_{RL,i}^{2}+\sigma_{RQ,i}^{2})}}{N},
\end{equation}
where N is the number of pixels in each stacked spectrum.
Having done this 1000 times, we obtain a $\chi^2_{\nu}$ distribution under the null hypothesis for each redshift range. By fitting this distribution with its parametric form, we can estimate the probability of obtaining our observed results under the null hypothesis.  

The results of the null hypothesis test for this 2D matched sample are presented in the upper left corner of each panel in Figure \ref{fig:RL_vs_RQ} and in Table \ref{tab:NH test}. For all redshift bins, we find $p$-values smaller than a significance level of $\alpha=0.05$, indicating that the two samples are not drawn from the same parent population. 

\begin{table*}
    \centering
    \caption{The $p$-values obtained from the null hypothesis test. The columns represent the results obtained from comparing: the RL and RQ populations with the 2D and 3D matched samples, the more radio-loud bins ($R_{2}$, $R_{3}$ and $R_{4}$) and the radio-quietest sample ($R_{1}$), and the radio-detected ($R_{1\rm{D}}$) and the radio-undetected ($R_{1\rm{U}}$) sample for the radio-quietest bin defined in section \ref{sec:radio bins} per a given redshift bin.}
    \begin{tabular}{|c|c|c|c|c|c|c|}
        \hline
        Redshift Range & (RL -  RQ)$_{\rm{2D}}$ & (RL - RQ)$_{\rm{3D}}$  & $R_{1}$ - $R_{2}$ & $R_{1}$ - $R_{3}$ & $R_{1}$ - $R_{4}$ &  $R_{1\rm{U}}$ - $R_{\rm{1D}}$ \\
        \hline
        0.6--0.8 & 0.01 & 0.03 & \multirow{2}{*}{$5.51\times 10^{-41}$}& \multirow{2}{*}{$7.90\times 10^{-15}$} & \multirow{2}{*}{$2.26\times 10^{-6}$}& \multirow{2}{*}{0.01}\\
        0.8--1.1 & $3.39\times 10^{-18}$ & $1.48\times 10^{-14}$ & \\
        \hline
        1.1--1.4 &  $2.31\times 10^{-53}$ & $3.014\times 10^{-51}$ & \multirow{2}{*}{$4.73\times 10^{-45}$} & \multirow{2}{*}{$1.43\times 10^{-18}$} &\multirow{2}{*}{$2.16\times 10^{-18}$} & \multirow{2}{*}{0.31}\\
        1.4--1.7 & $2.83\times 10^{-26}$ & $1.88\times 10^{-25}$ & \\
        \hline
        1.7--1.9 & $2.88\times 10^{-45}$ & $3.40\times 10^{-28}$& \multirow{2}{*}{$2.38\times 10^{-38}$}& \multirow{2}{*}{$1.92\times 10^{-23}$} &\multirow{2}{*}{$1.37\times 10^{-17}$}  & \multirow{2}{*}{ 0.05}\\
        1.9--2.1 & $7.43\times 10^{-26}$ & $1.67\times 10^{-13}$ & \\
        \hline
        2.1--2.4 & $1.03\times 10^{-39}$ & $5.15\times 10^{-24}$& \multirow{2}{*}{$6.61\times 10^{-30}$}& \multirow{2}{*}{$4.59\times 10^{-29}$} &\multirow{2}{*}{$1.25\times 10^{-11}$} & \multirow{2}{*}{0.08} \\
        2.4--2.7 & $1.45\times 10^{-36}$ & $2.02\times 10^{-24}$ & \\
        \hline
        2.7--3.0 & $3.21 \times 10^{-24}$ & $6.22\times 10^{-23}$& \multirow{2}{*}{0.001}& \multirow{2}{*}{$3.28 \times 10^{-9}$} &\multirow{2}{*}{$5.95 \times 10^{-5}$} & \multirow{2}{*}{0.86} \\
        2.7--3.4 & $3.90 \times 10^{-16}$ & $1.38 \times 10^{-10}$ & \\
        \hline
    \end{tabular}
    \label{tab:NH test}
\end{table*}

\subsection{Black Hole Mass Dependence}\label{sec:MBH}

\begin{figure*}
\includegraphics[width=\textwidth]{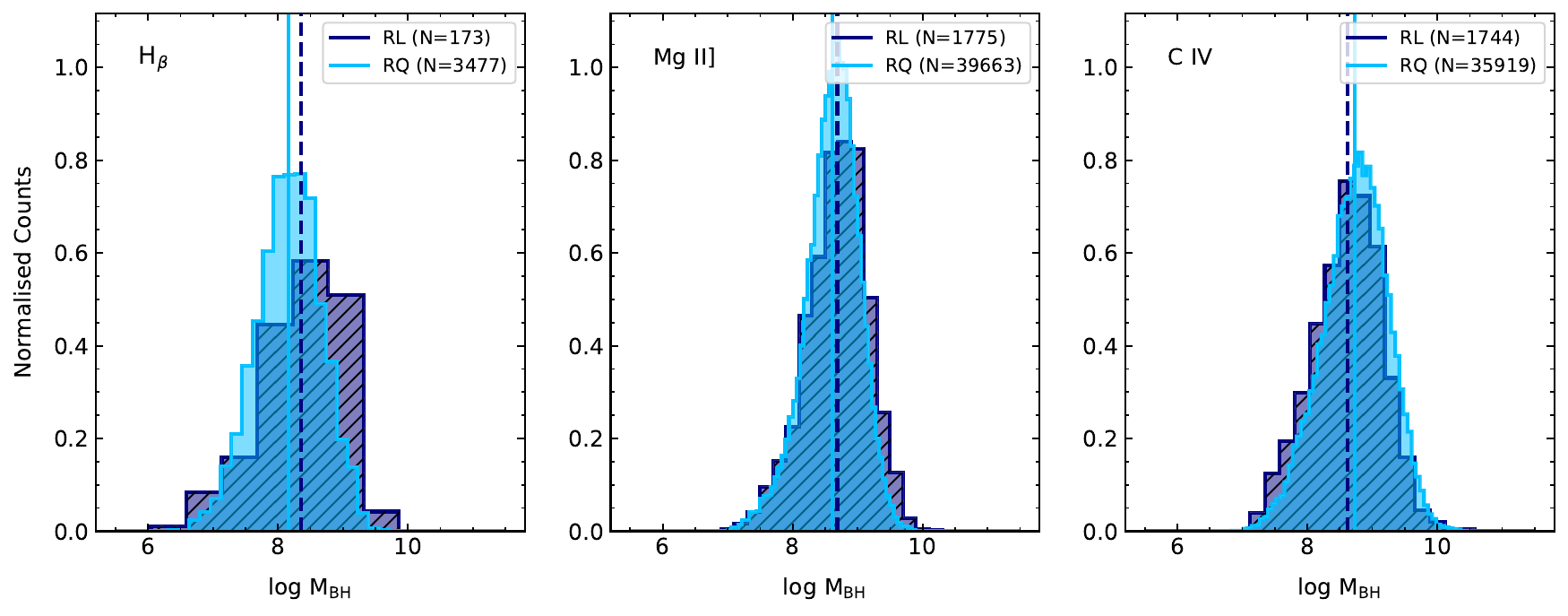}
    \caption{The black hole mass estimates provided from \citet{rakshit2020spectral} for the radio-loud (dark blue) and radio-quiet population matched in redshift and $i$-band magnitude (light blue). Each panel corresponds to the three different methods of calculation, which depend on the availability of a given emission line: the H$_{\beta}$ line (z<0.8), the Mg II line (0.8$\leq z$ < 1.9) and the C IV line ($z\geq$ 1.9). The sample means for the RL and RQ QSOs are presented in each panel as dashed dark blue and solid blue lines, respectively.}
    \label{fig:MBH_hist}
\end{figure*}

As discussed in the introduction, several studies have found the black hole mass to be a defining factor in the radio-loudness dichotomy, where RL QSOs are found to harbour black holes with $M_{\rm{BH}}\gtrsim 10^{8}$ (e.g. \citealt{dunlop2003quasars}, \citealt{mclure2004relationship}; \citealt{chiaberge2011origin}). This, however, does not appear to be the case for our sample. In Figure \ref{fig:MBH_hist} we present the black hole mass distributions for each radio class using the virial black hole mass estimates from \cite{rakshit2020spectral}. We can see that both the RL (dark blue) and RQ QSOs (light blue) are found to span similar $M_{\rm{BH}}$ ranges, irrespective of the different black hole mass calibrations. In addition, the mean black hole masses do not show any systematic trend - the RL appear more massive for the H{\sc ${\beta}$} and Mg\,{\sc ii}, but less so for the C\,{\sc iv} calibration (cf. \citealt{mclure2004relationship}). However, to determine whether this parameter plays a role in causing the differences observed in Figure \ref{fig:RL_vs_RQ}, we re-do the stacking procedure with the $z-M_{i}-M_{\rm{BH}}$ sample, i.e. including the M$_{\rm{BH}}$ in the process, even though it reduces the sample size. We note that by matching in $M_{i}$ and $M_{\rm{BH}}$ that we are also effectively controlling for the Eddington ratio.

The resulting RL/RQ comparison is found to exhibit similar features as for the $z-M_{i}$ sample, where both the continuum and the emission lines are found to differ between the two populations (see Appendix \ref{sec:supplementary} for the RL/RQ comparison with the $z-M_{i}-M_{\rm{BH}}$ sample). We use the null hypothesis test as described in the previous section to determine the significance of this result. Similarly to the $z-M_{i}$ sample, we are able to rule out the null hypothesis with $p<0.05$ for all redshift bins (see results for the 3D matched sample in Table \ref{tab:NH test}). Therefore, we conclude that additional information beyond the black-hole mass and accretion rate is required to explain the difference between the RL and RQ population. As matching in $M_{\rm{BH}}$ is on average of little consequence (except for sample size), we proceed with the rest of our analysis using only the $z-M_{i}$ sample, since it is about three times larger. 

\subsection{Comparison as a Function of Radio-loudness}\label{sec:radio bins}
 
\begin{figure*}
\includegraphics[width=\textwidth]{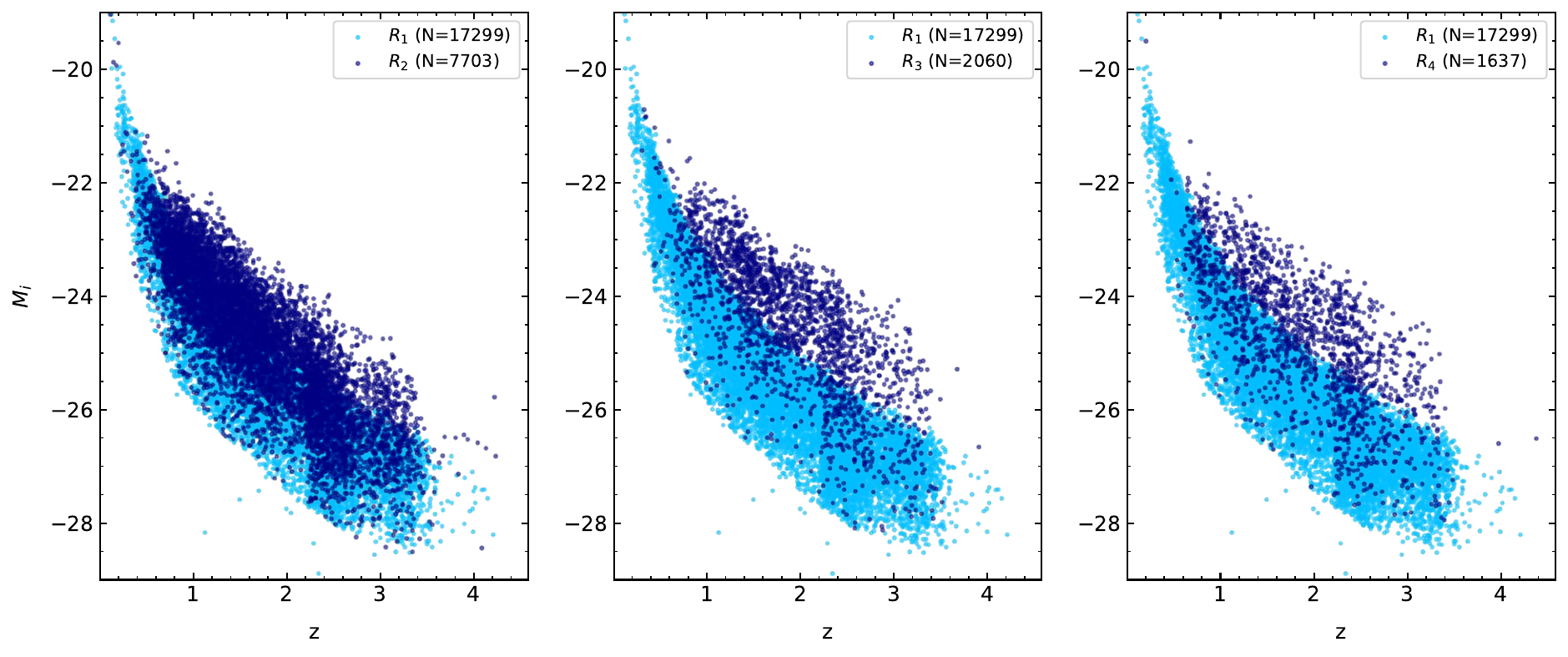}
    \caption{The $z-M_{i}$ plane for the different radio-loudness bins. The lowest radio-loudness bin ($R_{1}$) is shown in light blue, whereas the bins with increasing R are presented in dark blue throughout the three panels.} 
    \label{fig:z_Mi}
\end{figure*}

To investigate whether the differences observed between the RL and RQ population represent a bimodality or a transition between populations, we further separate our sample into four radio-loudness bins as indicated in Figure \ref{fig:R}. As with our previous approach, we include radio-undetected sources only in the lowest radio-loudness bin ($R_{1}$) as we cannot confidently assign the rest of the sources to the correct $R$ bin (i.e when the upper limits fall to the right of the demarcation lines in Figure \ref{fig:R}).  
This leads to some challenges when matching in $z$ and $M_{i}$ as demonstrated by Figure \ref{fig:z_Mi}, which presents the redshift-luminosity plane. We can see that the more radio-loud bins ($R_{2}$, $R_{3}$ and $R_{4}$) in comparison to $R_{1}$ span a different parameter space. This means that we need to match to the lowest radio-loudness bin sample, unlike before where the matching was performed to the RL population in order to maximise the radio-loudest sample. As this resulted in a lower number of sources for each radio-loudness bin, we decided to reduce the number of redshift bins by doubling the bin width to maintain comparable signal-to-noise levels of the stacked spectra as before.

The results of applying our spectral stacking algorithm and the null hypothesis test on this sample division for five redshift bins show that all radio-loud bins exhibit statistically significant differences when compared to $R_{1}$ (see results for $R_{1}$-$R_{2}/R_{3}/R_{4}$ in Table \ref{tab:NH test} and Appendix \ref{sec:supplementary}), providing no indication of a smooth transition between populations relative to the null hypothesis test.To address the possibility of a bias arising from our selection process, where only radio-detected sources were included in the more radio-loud bins, we further conducted a comparison between radio-detected ($R_{\rm{1D}}$) and radio-undetected sources ($R_{\rm{1U}}$) within the lowest radio-loudness bin ($R_{1}$). We find that we can accept the null hypothesis that the two samples are drawn from the same parent distribution ($p\geq0.05$) for all, apart from the lowest redshift bin, where $p=0.01$ indicating ambiguity. This ambiguous result, however, is not enough to explain $p$-values as low as $10^{-41}$ as found for the $R_{1}$-$R_{2}$ comparison. Therefore, we conclude that the exclusion of non-detections is not significantly impacting our findings. 

We must also consider the possible role of the host galaxy (e.g. \citealt{magliocchetti2020role}). Given our high S/N, we expect to be able to detect starlight up to redshifts normally inaccessible to SDSS; light from the hosts could therefore be contributing to our results. To test this, we examine the excess region of flux density for each pair per given redshift bin. Our findings indicate that the luminosity of this region is of order $10^{9-10}L_{\odot}$, suggesting that the host galaxy might indeed play a significant role in generating the observed effects.

Another possible explanation for the absence of a smooth transition between populations in terms of the hypothesis test results is that we are searching for differences only in the average spectrum in each bin. Since \cite{macfarlane2021radio} showed that there exists a significant population of jetted QSOs even at the lowest $R$ values, it is possible that the differences we see are a result of an increasing fraction of jetted AGN which become numerically dominant in all but the lowest $R$ bin, and which therefore dominate the median stacked spectra. This hypothesis will become testable in the future as we are increasingly able to morphologically discern the presence of jets in forthcoming wide field sub-arcsecond 144 MHz imaging from the LOFAR international stations.

\section{Spectral Properties of RL and RQ quasars}\label{sec:spec_prop}

In this section, we employ the \textsc{\texttt{PyQSOFit}} spectral fitting code to obtain spectral properties from the quasar populations. In section \ref{sec:em}, we analyse the average emission line properties of both RL and RQ quasars, as well as those categorized into different radio-loudness bins, while in section \ref{sec:OII}, we focus on investigating the nature of the [O\,{\sc ii}] excess found for the RL regime.

\subsection{The Spectral Fitting Procedure}\label{sec:fitting}
To investigate the continuum and emission line properties of the composite spectra of radio-loud and radio-quiet quasars we use the \textsc{\texttt{PyQSOFit}} spectral fitting code in a similar manner to previous works (e.g. \citealt{shen2019sloan}, \citealt{rakshit2020spectral}, \citealt{fawcett2022fundamental}), where the continuum is fitted globally (i.e. the regions influenced by the presence of emission lines are masked out), whereas the emission line complexes are fitted separately and locally. 

The continuum of each composite spectrum is globally modelled by using a combination of power-law, Balmer continuum, Fe\,{\sc ii} component and third-order polynomial as:
\begin{equation}
    f_{\rm{conti}}=f_{\rm{pl}}+f_{\rm{BC}}+f_{\rm{Fe\,\textsc{ii}}}+f_{\rm{poly}}.
\end{equation}

The power-law continuum component has been added to represent the emission from the accretion disc and is defined as:
\begin{equation}
f_{\rm{pl}}=a_{0}(\lambda/\lambda_{0})^{a_{1}},
\end{equation}
where $a_{0}$ and $a_{1}$ are the normalisation and power-law slope, and $\lambda_{0}$ is a reference wavelength at 3000 $\angstrom$.

The Balmer continuum component represents the sum of blended, higher-order Balmer lines that give rise to the well-known small blue bump at $\lambda\sim3000\angstrom$. In \textsc{\texttt{PyQSOFit}} it is modelled by the function given by \cite{grandi19823000} for the case of optically thick clouds as:
\begin{equation}
    f_{\rm{BC}} = F_{\rm{BE}} B_{\lambda}(T_{e}) (1-e^{-\tau_{\lambda}}(\lambda/\lambda_{\rm{BE}})^{3}),
\end{equation}
where $\lambda_{\rm{BE}}$ is the position of the Balmer edge, $F_{\rm{BE}}$ is the flux at the Balmer edge, $B_{\lambda}(T_{e})$ is the Planck function at the electron temperature $T_{e}$ and $\tau_{\lambda}$ is the optical depth. Here, $F_{\rm{BE}}$,  $T_{e}$ and $\tau_{\lambda}$ are left as free parameters as in \cite{fawcett2022fundamental}.

The Fe\,{\sc ii} component is another essential part of modelling the continuum. At UV and optical wavelengths, AGN spectra contain numerous iron emission lines that blend together to form a pseudo-continuum. If not properly subtracted, this pseudo-continuum could contaminate the continuum and emission line measurements. Here, \textsc{\texttt{PyQSOFit}} models it as:
\begin{equation}
    f_{\rm{Fe\,\textsc{ii}}}=c_{0}F_{\rm{Fe\,\textsc{ii}}}(\lambda,c_{1},c_{2}),
\end{equation}
where $c_{0}$ is the normalisation constant, $c_{1}$ is a constant describing the Gaussian broadening and $c_{2}$ represents the wavelength shift applied to the Fe\,{\sc ii} templates to match the data. For the UV part of the spectrum, we use the modified UV Fe\,{\sc ii} template by \cite{shen2019sloan} that combines the \cite{vestergaard2001empirical} template for rest-frame wavelengths 1000–2200 $\angstrom$, the \cite{salviander2007black} template for 2200–3090 $\angstrom$, and the \cite{tsuzuki2006fe} template for 3090–3500 $\angstrom$. The optical Fe\,{\sc ii} template, on the other hand, is taken from \cite{veron2004unusual}, covering the rest-frame wavelengths of 3535–7534 $\angstrom$.

Finally, the third-order polynomial is used to account for the bending of the continuum, which is likely caused by the intrinsic dust reddening of the population, and is defined as:
\begin{equation}
    f_{\rm{poly}}=\sum^{i=3}_{i=0}b_{i}(\lambda-\lambda_{0})^{i},
\end{equation}
where $b_{i}$ are the model free parameters.

After subtracting the best-fit model of the continuum, we fit the emission line spectrum in log-space. The fitting is performed individually for each emission line complex, where all the emission lines contained within a single complex are fit simultaneously. The narrow components (FWHM < 1200 km s$^{-1}$) are modelled by a single Gaussian, where the velocity offset and line width are tied for each line complex. The broad emission line profiles, however, can be quite complex (e.g., double-peaked, with a flat top, or asymmetric) and thus are not well represented by a single Gaussian. In such cases, we use multiple Gaussian components depending on the line complexity. A full list of the line complexes containing the individual emission lines and number of Gaussian components used in the fit is provided in Table \ref{tab:EM_lines}.

From the best-fitting models, we can obtain emission line fluxes for each line in question. The uncertainties of these measurements are calculated using the Monte Carlo approach embedded within PyQSOFit, where we choose to use 100 iterations. 

\begin{table}
    \centering
    \caption{The emission line fitting information. The columns are as follows: the name of the line complexes, the wavelength range used for the fit, the emission lines present in each line complex and the number of Gaussian components used for each line.}
    \begin{tabular}{c|c|c|c}
    \hline
         Emission Line &  Wavelength & Emission  & Number of\\
         Complex&    Range  & Line&   Gaussian   \\
         Name&  ($\angstrom$)& Name &  components\\
         \hline
         H{\sc${\beta}$}&4640-5100&H{\sc${\beta}$} broad&2\\
         &&H{\sc${\beta}$} narrow&1\\
         & & $[$O\,{\sc III}$]$ (4959 $\angstrom$)&1\\
         & & $[$O\,{\sc III}$]$ (5007 $\angstrom$)&1\\
         H{\sc ${\gamma}$}&4200-4440&H{\sc${\gamma}$} broad&1\\
         & & H{\sc $_{\gamma}$} narrow&1\\
         & & O III (4363 $\angstrom$)&1\\
         $[$O\,{\sc ii}$]$& 3650-3800& $[$O\,{\sc ii}$]$ (3728 $\angstrom$)&1\\
         $[$Ne\,{\sc v}$]$&3380-3480&$[$Ne\,{\sc v}$]$ (3426 $\angstrom$)&1\\
         Mg\,{\sc ii]}&2700-2900& Mg\,{\sc ii]} broad&2\\
         && Mg\,{\sc ii]} narrow&1\\
         C\,{\sc iii]}&1810-1970&C\,{\sc iii]} broad&2\\
         && C\,{\sc iii]} narrow&1\\
         C\,{\sc iv}&1500-1700&C\,{\sc iv} broad&2\\
         &&C\,{\sc iv} narrow&1\\
         &&He\,{\sc ii} (1640 $\angstrom$) broad&1\\
         &&He\,{\sc ii} (1640 $\angstrom$) narrow&1\\
         &&O\, {\sc iii]} (1663 $\angstrom$) broad&1\\
         &&O\,{\sc iii]} (1663 $\angstrom$) narrow&1\\
         Si\,{\sc iv} + O\,{\sc iv]} &1290-1450&Si\,{\sc iv} + O\,{\sc iv]} \\
         Ly{\sc $\alpha$}&1150-1290&Ly$\alpha$ broad&2\\
         &&Ly{\sc $\alpha$} narrow&1\\
         &&N\,{\sc v} (1240 $\angstrom$)&1\\
         O\,{\sc vi} &980-1100&O\,{\sc vi} &1\\
         \hline
    \end{tabular}
    \label{tab:EM_lines}
\end{table}

\subsection{Average Emission Line Properties}\label{sec:em}

As mentioned in section \ref{sec:SDSS}, \textsc{\texttt{PyQSOFit}} contains an additional feature that separates a given quasar spectrum into host galaxy and pure quasar contribution. However, for this separation to be well represented, the continuum needs to be accurately fitted. 
Unfortunately, the host and QSO continuum models implemented in \textsc{\texttt{PyQSOFit}} prove inadequate for our objectives, as demonstrated by poor fitting statistics (see example fit in Appendix \ref{sec:supplementary}). They appear to be strongly dependent on the details of the wavelength range used, indicative of a local minimum problem. Moreover, these high $\chi_{\nu}^{2}$ values have additional implications as they prevent us from applying dust-extinction laws to investigate the underlying cause of the continuum differences observed in section \ref{sec:comparison}. We are, however, able to obtain acceptable fits for the emission lines. To improve their goodness-of-fit, we use an additional localised power-law continuum component for each line complex after subtracting the globally fitted continuum, following the method of \cite{fawcett2022fundamental}. In addition, we perform a flux calibration before the fit to obtain non-arbitrary units. This is done by shifting each composite spectrum to the observed-frame by using its central redshift value ($z_{\rm{centre}}$) and the SDSS $i$-band filter curve and the median apparent $i$-band magnitude of the individual spectra to obtain a flux density in units of erg s$^{-1}$ cm$^{-2}$ $\angstrom^{-1}$.

\begin{figure*}
\includegraphics[width=\textwidth]{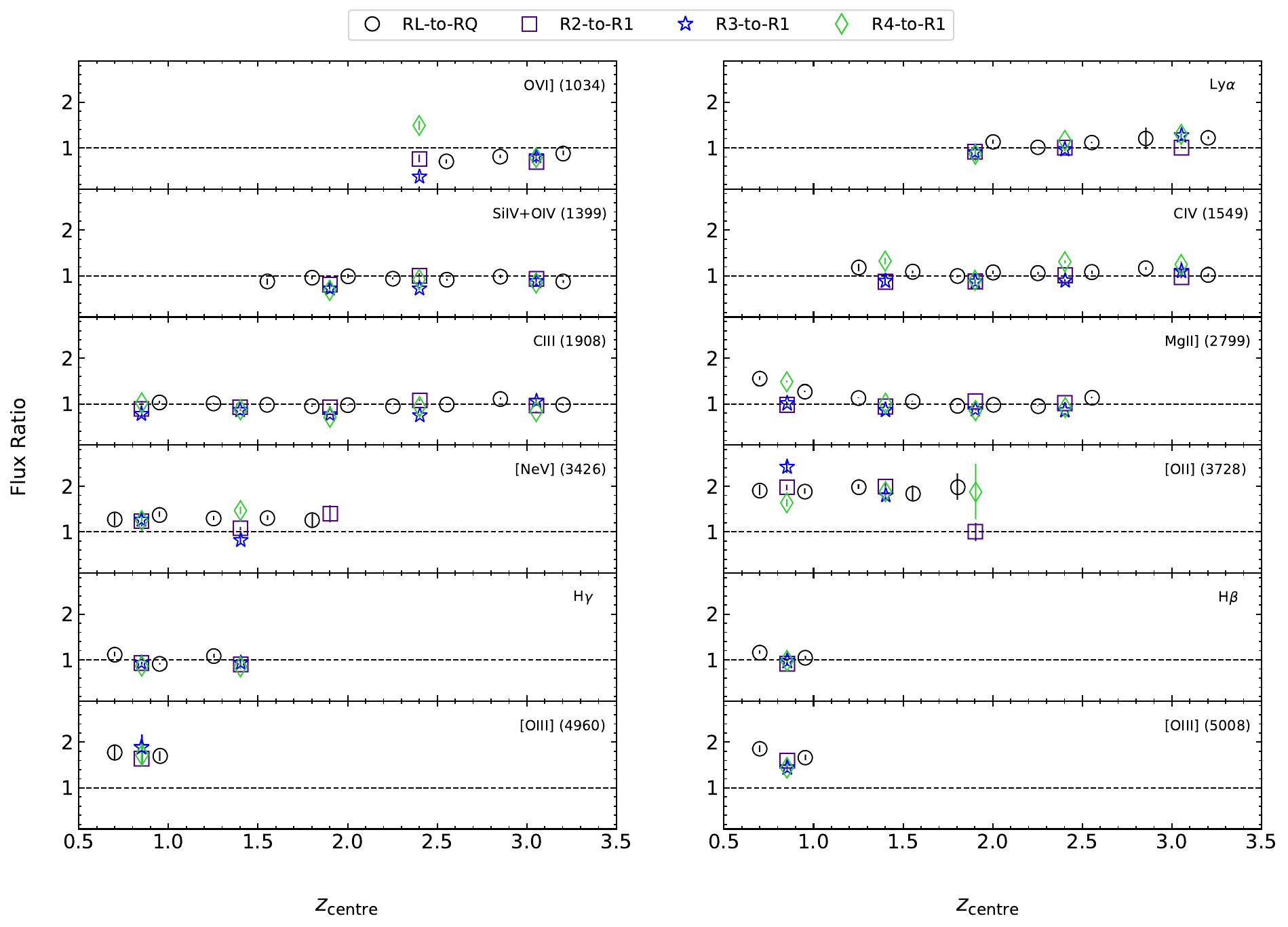}
    \caption{ The emission line flux ratio of a given pair, as indicated in the legend, as a function of redshift. Each panel represent an emission line under comparison as denoted in the upper right corner, which must be significantly detected (>3$\sigma$) for inclusion. The dashed line in each panel represents the line of equality, facilitating visual comparison of the emission line flux ratios between the different pairs.}
    \label{fig:EM_lines}
\end{figure*}

Figure \ref{fig:EM_lines} presents the ratio of the emission line flux of the RL to RQ QSOs (RL-RQ), along with the more radio-loud to radio-quietest bin ($R_{2}$/$R_{3}$/$R_{4}$-$R_{1}$) for the most prominent emission lines. Across all considered pairs, we find that the majority of emission lines exhibit flux ratios close to unity. However, there is a strong excess for radio-loud objects in the [Ne\,{\sc v}] 3426 $\angstrom$, [O\,{\sc ii}] 3728 $\angstrom$ and the [O\,{\sc iii}] 4960/5008 $\angstrom$ (hereafter [Ne\,{\sc v}], [O\,{\sc ii}] and [O\,{\sc iii}]) emission lines. These lines are of particular interest as [Ne\,{\sc v}] is used to trace extended emission line regions (EELR: see discussion in next section), whereas the [O\,{\sc iii}] emission line is generally associated with the AGN bolometric luminosity (\citealt{heckman2004present}; \citealt{best2012fundamental}; \citealt{kalfountzou2012star}), as it is mostly ionised by the continuous radiation coming from the accretion disc. The [O\,{\sc ii}] emission line, on the other hand, is a well-known star formation tracer in quasar host galaxies (e.g. \citealt{ho2005ii}; \citealt{kalfountzou2012star}; \citealt{matsuoka2015sloan}) as in contrast with the [O\,{\sc iii}], it is only weakly excited in the narrow line region (NLR). However, as result of the low number of data points for the [O\,{\sc iii}] emission due to the available redshift range, we focus only on [Ne\,{\sc v}] and [O\,{\sc ii}] in the following section.

\subsection{Star Formation Rates}\label{sec:OII}

The [O\,{\sc ii}] forbidden line is a prominent feature that can be easily detected in moderate resolution spectra, making it a good SF tracer in the absence of H$\alpha$. When an AGN is present, however, this emission line can be severely contaminated by excitation from EELR which can span out to several kpc (e.g. \citealt{unger1987extended}; \citealt{villar2011interactions}; \citealt{husemann2014integral}; \citealt{maddox2018ii}). Therefore, to determine whether the elevated levels of [O\,{\sc ii}] found in section \ref{sec:em} are due to SF, we employ the AGN correction technique described in \cite{maddox2018ii}. Briefly, this method involves removing sources from the stacks with a [Ne\,{\sc v}] detection and making use of the [O\,{\sc ii}]-to-[Ne\,{\sc v}] emission line flux ratio, which in a typical AGN-dominated galaxy is found to be of order unity as a result of excitation from NLR. The [Ne\,{\sc v}] is chosen because of its wavelength proximity to [O\,{\sc ii}] and its high ionisation potential, making the effects of dust attenuation and star-forming regions negligible. 

As the spectral measurements provided by \cite{rakshit2020spectral} did not contain any information regarding [Ne\,{\sc v}], we used \textsc{\texttt{PyQSOFit}} to fit a single Gaussian in a spectral window of 150 $\angstrom$ centered on the [Ne\,{\sc v}] central wavelength. The emission line is considered as a non-detection if the measured emission line flux is detected at a S/N<3. Although in previous sections we investigated how the optical properties varied in bins of radio-loudness, the need to remove the [Ne\,{\sc v}] detections from the stacks (which involves about 20 per cent) reduces the sample size to a point that it is no longer possible to search for evolution in both redshift and radio-loudness. Therefore, we continue the rest of the analysis with the RL and RQ populations. However, we also create a new sample matched in $z$ and $M_{i}$, with the division set at $R=0.2$ (i.e. the separation between $R_{1}$ and the more radio-loud bins), to check whether using a slightly different classification would yield different results.
From the newly computed composite spectra, we obtained emission line measurements for [O\,{\sc ii}] and [Ne\,{\sc v}] in the same manner as before. A [Ne\,{\sc v}] detection is still present in the stacks as a result of the high S/N which is sufficient to detect the contribution from the NLR. Using these measurements and the value of [O\,{\sc ii}] / [Ne\,{\sc v}] = 1.05 obtained from the SDSS composite spectrum from \cite{berk2001composite}, we are able to derive a measure of the [O\,{\sc ii}] emission line flux unaffected by AGN. 

\begin{figure}
\includegraphics[width=\columnwidth]{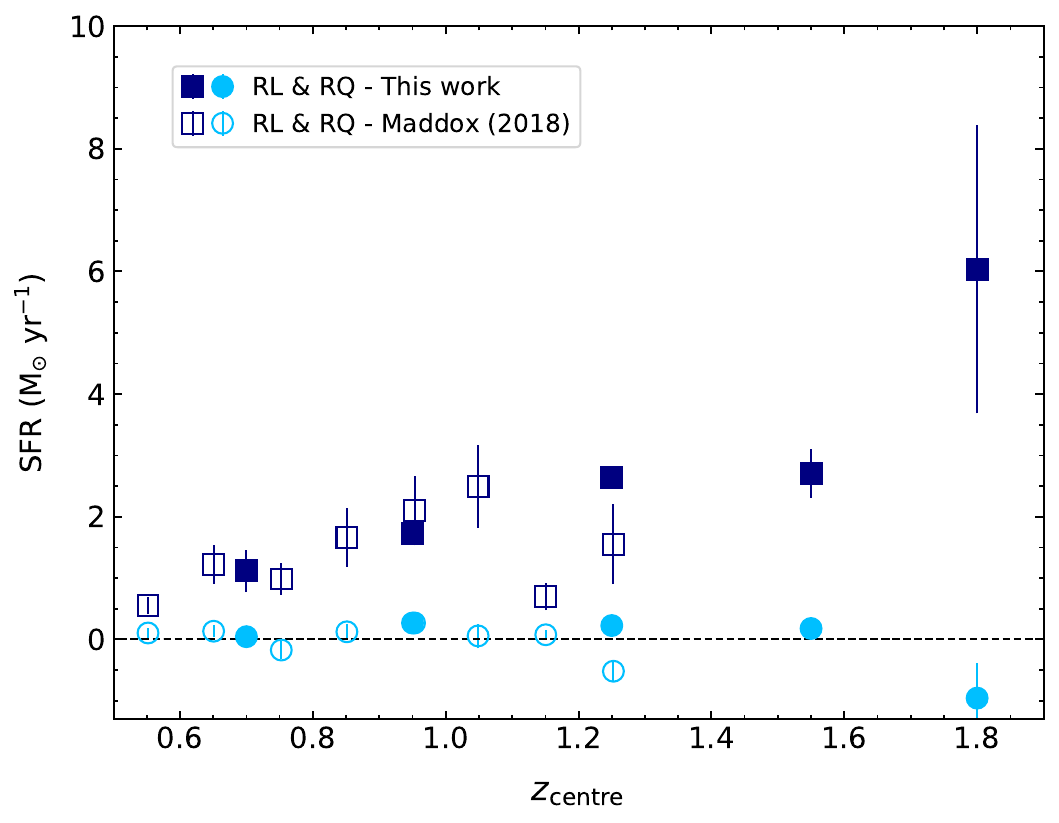}
\caption{The SFR obtained from the doubly corrected [O\,{\sc ii}] emission line flux for the RL (dark blue) and RQ population (light blue) compared against the SFR
obtained from \citet{maddox2018ii}. The black dashed line indicates a SFR of zero.}
\label{fig:SFR}
\end{figure}

Figure \ref{fig:SFR} shows the [O\,{\sc ii}] derived, doubly corrected SFR against redshift obtained for the RL and RQ populations using the [O\,{\sc ii}] - SFR conversion from \cite{kewley2004ii}. We can see that the SFR of the RQ population is close to zero, indicating that on average star-formation processes in the host galaxy of RQ QSOs are not sufficient to produce a strong [O\,{\sc ii}] emission line. In contrast, the excess of [O\,{\sc ii}] persists for the RL population even when correcting for contamination from EELR and NLR. The alternative classification ($R>0.2$) also exhibits the same trend, further indicating that the exact value of $R$ is not important.
Our results are consistent with those of \cite{maddox2018ii} who used SDSS DR7 with Faint Images of the Radio Sky at Twenty-cm (FIRST; \citealt{becker1995first}), but we have shown that their findings persist with higher statistical significance and to a further $\approx 1.4$\,Gyr of cosmic history, closer to the peak epoch of cosmic star formation rate density \citep{madau2014cosmic}. This is because we benefit not only from the increased sensitivity of LoTSS and the reduced impact of line of sight effects (relative to the 1.4\,GHz data), but also the extended wavelength coverage of the BOSS spectrograph \citep{smee2013multi}.

\section{Discussion}\label{sec:discussion}

\subsection{Could our results be caused by starburst galaxies contaminating our RL sample?}\label{sec:contaminants}

As discussed in the introduction, \cite{macfarlane2021radio} have shown that the radio flux density of quasars can be modelled by a combination of star formation and radio jets. This naturally leads to the question of whether the [O\,{\sc ii}] excess found in section \ref{sec:OII} could be caused by having radio-quiet quasars hosted by starburst galaxies in our RL sample. The presence of intense star formation in such hosts can lead to a larger degree of reddening (\citealt{gordon1997dust}), which would further explain the redder continuum observed in section \ref{sec:comparison}. In addition, this scenario does not require any differences between the black hole mass and accretion rate of the QSOs themselves (i.e. $R_{\rm{edd}}$), which is in agreement with our results (see section \ref{sec:MBH}). 

To explore this possibility, we used the main-sequence relation from \cite{schreiber2015herschel} and an offset of 0.6 dex based on the starburst criterion from \cite{rodighiero2011lesser} to obtain a SFR value for a typical starburst galaxy with a stellar mass of $10^{11}$ M$_\odot$ at a redshift equal to our central bin values from Figure \ref{fig:SFR}. This resulted in SFR values ranging from 127 - 488 M$_\odot$ yr$^{-1}$ for the redshift range of $0.6<z<1.9$. These estimates are then converted to a radio luminosity at 150 MHz by using the mass independent SFR-$L_{150\rm{MHz}}$ relation from \cite{smith2021lofar}. Comparing these results to the RL sample, we find that up to 5 per cent in any given redshift bin, could have a radio luminosity consistent with contribution from RQ starbursts. However, taking into account the large difference between our doubly corrected [O\,{\sc ii}] SFR and the values obtained here for the starburst galaxies, it is unlikely that the radio emission is due to such extreme star-forming processes. Furthermore, considering the fact that we are investigating the median stacks of the populations, we believe that if starburst galaxies are present in our samples they will not noticeably impact our results.

\subsection{Could shock excitation be playing a role?}\label{sec:shocks}

In section \ref{sec:OII}, we have used the [Ne\,{\sc v}] emission line to remove contamination from EELR and correct for contribution from the AGN NLR. However, shocks in the interstellar medium generated by radio jets or AGN-driven winds could also contribute to the [O\,{\sc ii}] emission. To assess their potential influence, \cite{maddox2018ii} used the \textsc{\texttt{MAPPINGS III}} shock and photoionization modelling code from \cite{allen2008mappings}. The author found that there is a variety of shock conditions capable of exciting [O\,{\sc ii}], but only high velocity shocks (>600 km s$^{-1}$) are able to produce [Ne\,{\sc v}]. This suggests that the doubly corrected [O\,{\sc ii}] could still be contaminated from moderate-velocity shocks. A prevalence of such shocks for the RL population may be able to explain the apparent enhancement of SFR and their redder appearance. 

\begin{figure}
\includegraphics[width=1.\columnwidth]{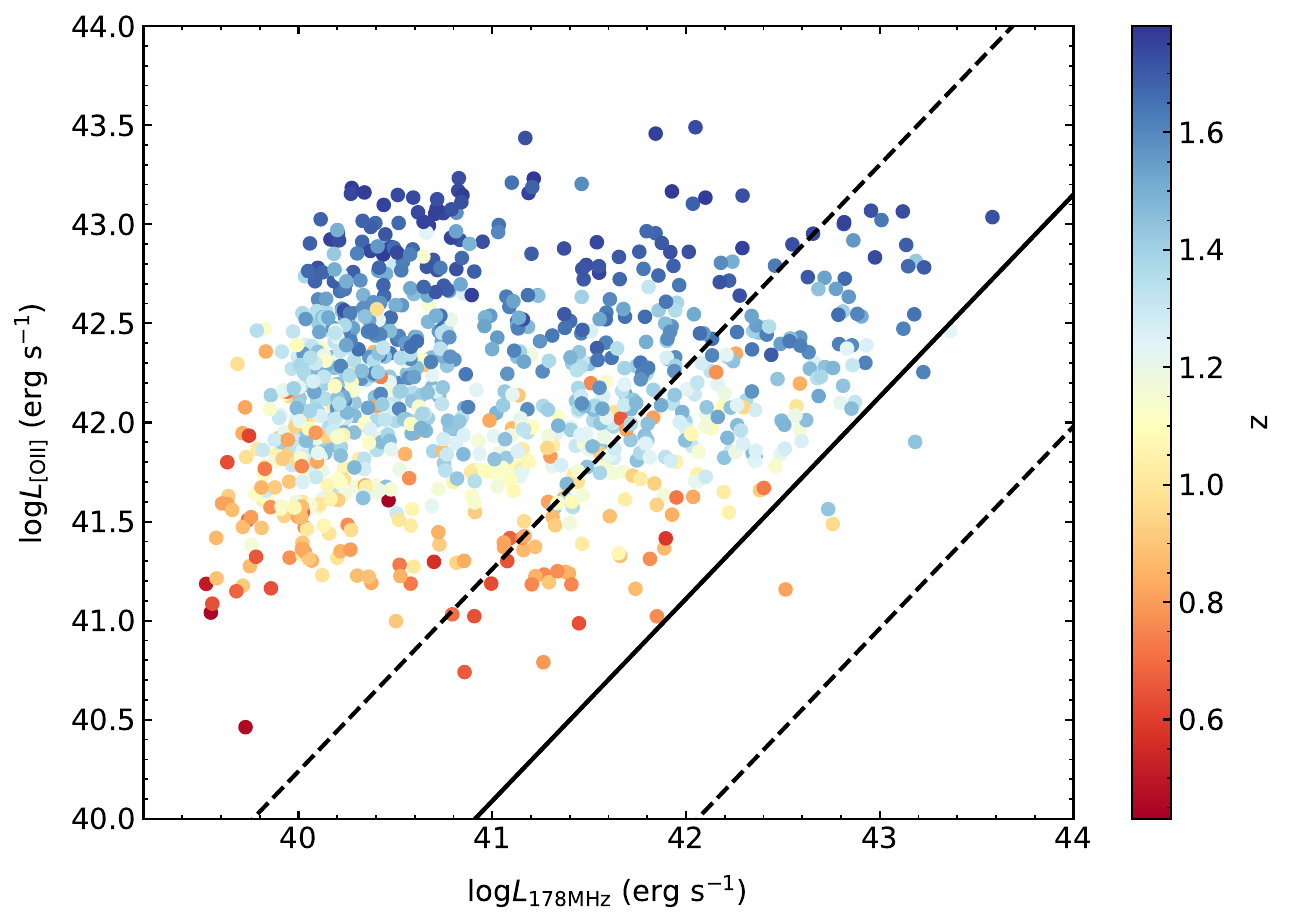}
\caption{The [O\,{\sc ii}] emission line luminosity as a function of 178MHz total radio luminosity for RL sources without significant [Ne\,{\sc v}] detections. The solid and dashed black lines are generated from the regression analysis results for the 3CRR radio sources in Table 6 of \citet{hardcastle2009active}, where the solid line indicates the regression line, whereas the dashed lines show $\pm3\times$ the obtained scatter.}
\label{fig:OII vs radio}
\end{figure}

To investigate this scenario, we explore the connection between [O\,{\sc ii}] emission and 178 MHz luminosity, where we use the low frequency radio emission as a proxy for jet power. Here, we use a radio spectral index of $\alpha=-0.7$ to convert from 144 MHz to 178 MHz, in order to compare our results with \cite{hardcastle2009active}, who have found a positive correlation for a sample of 3CRR radio sources, which is thought to be of nuclear origin. The results for the RL sample from section \ref{sec:OII}, which was used to calculate the [O\,{\sc ii}] SFRs, are presented in Figure \ref{fig:OII vs radio}. We can see that the [O\,{\sc ii}] appears to be independent of the radio emission at any given redshift. However, there is a subset of sources that lie within the region defined by the 3CRR sample. This indicates the presence of some AGN-related influence, which can be as high as 36 per cent for the lowest redshift range ($0.6<z<0.8$), and up to 13 per cent for the highest range ($1.7 < z < 1.9$). But, as previously discussed our analysis relies on SFR derived from the median composite spectra of the populations. Furthermore, the correction for the AGN NLR is not factored into the individual RL sources, which could potentially explain the correlation observed in the 3CRR sources. Therefore, if shocks are present in the RL population, we believe that they alone cannot account for the observed differences between RL and RQ quasars.

\subsection{Possible explanations}\label{sec:explanation}

To explain the differences between the RL and RQ population, we propose two distinct models: one centered on black hole spin dynamics (\citealt{blandford1977electromagnetic}; \citealt{wilson1994difference}; \citealt{sikora2007radio}) and another on the evolutionary scenario proposed for red and blue quasars (e.g. \citealt{sanders1990ultraluminous}; \citealt{hopkins2008cosmological}; \citealt{alexander2012drives}; \citealt{klindt2019fundamental}; \citealt{fawcett2020fundamental, fawcett2022fundamental}). 

The first model involves a rapidly spinning black hole, coupled with a rich gas supply required for accretion and jet production (e.g. \citealt{hardcastle2007hot}; \citealt{gurkan2015herschel}). Given that the gas needed for accretion, which leads to the radio-loudness of the source, is also essential for fueling star formation, these two processes are inherently interconnected. Therefore, the question becomes how to get a rapidly spinning black hole and a rich gas supply to make a RL quasar. We suggest two plausible scenarios: Firstly, the traditional major merger event which results in a rapid inflow of gas and dust. This material not only transports angular momentum toward the central region, leading to the rapid spin-up of the black hole, but also serves as a trigger for star formation. Secondly, we can consider a scenario in which RL quasars are hosted by massive, gas-rich spiral galaxies. In this setting, a substantial reservoir of cold gas is available, supporting continuous star formation and accretion. Furthermore, due to the ordered rotation characteristic of spiral galaxies, it may give rise to an efficient transfer of angular momentum onto the supermassive black hole, eliminating the necessity for merger events. 

The alternative model assumes once more a gas-rich merger triggering an AGN, but here we focus on different stages of evolution, following e.g. \citet{hopkins2008cosmological}. The first one is a relatively short-lived phase where the QSO is heavily obscured by high-column gas density and dust. Subsequently, the AGN generates powerful winds and/or outflows, which disperse the obscuring material. In our study, RL quasars could represent the former evolutionary stage, where the redder continuum and enhanced SFR would be explained by the obscuring and dense material, whereas RQ quasars may fall into the latter (unobscured) category. Here we do not require any difference in either the accretion rate or in the BH spins between the two classes.
This scheme is related to the one presented for red and blue QSOs discussed in previous studies (e.g. \citealt{klindt2019fundamental}; \citealt{fawcett2020fundamental}; \citealt{fawcett2022fundamental}), where dividing the QSO population according to their optical colours (rather than radio-loudness) gives a red class with a significant radio flux excess relative to the blue class. However, both the red and blue QSO classes contain RL and RQ sources, and the average QSO in both classes is consistent with being radio-quiet (i.e. $R < 1$). It is therefore clear that this association alone cannot explain our results (certainly we cannot equate the RL QSOs with the `red QSO' class, etc).

To make further progress on what controls the radio-loudness of QSOs, we need more information. This could come from the sub-arcsecond 144 MHz imaging that is now becoming possible with LOFAR (and the morphological information that it can provide; e.g. \citealt{morabito2022identifying}), along with larger statistical samples from new and forthcoming facilities such as WEAVE \citep{dalton2012weave,jin2023wide}, the Dark Energy Spectroscopic
Instrument (DESI; \citealt{aghamousa2016desiI, aghamousa2016desiII}) and the Multi-object Optical and Near-IR spectrograph (MOONs; \citealt{cirasuolo2014moons}). In addition, improved black hole spin estimates from X-ray observatories such as the European Space Agency's Athena X-ray observatory \citep{barcons2017athena} and NASA's Nuclear Spectroscopic Telescope Array (NuSTAR; \citealt{harrison2010nuclear}) will be crucial in deepening our understanding of this subject.

\section{Summary and Future Prospects}\label{sec:conclusion}

In this work we have used the second data release of the LOFAR Two-metre Sky Survey and the fourteenth data release of the Sloan Digital Sky Survey to create the largest, uniformly-selected, spectroscopically-confirmed sample of radio-loud and radio-quiet quasars. To study their spectroscopic properties, we have developed a new spectral stacking code which accounts for a range of biases including those that arise as a result of the redshifting process. Such a tool not only allows us to robustly compare quasar and galaxy populations, but also enables us to statistically recover the continuum properties of faint sources. This will become particularly important for radio-selected spectroscopic surveys such as the WEAVE-LOFAR survey \citep{smith2016weave}, which will generate more than one million optical spectra, with continuum detections absent in a significant fraction. Using this algorithm to investigate the average properties of QSOs as a function of their radio-loudness, we have found the following results:

\begin{itemize}
    \item[-] The high S/N composite spectra representative of the RL and RQ populations differ. RL QSOs are found to have on average a redder continuum with an [O\,{\sc ii}] emission line excess across the redshift range of $0.6<z<3.5$. Such differences highlight the importance of creating high-resolution stacks of both populations to improve the redshift classification of future spectroscopic surveys.
    
    \item[-] The RL and RQ population are found to span similar black hole mass ranges with no systematic trend showing that the mean $M_{\rm{BH}}$ is higher for RL QSOs. Furthermore, using a sample matched in $z$, $M_{i}$ and $M_{\rm{BH}}$ to make a comparison between the average spectra of RL and RQ QSOs is found to give similar statistically consistent results as for a sample matched in $z$ and $M_{i}$. This suggests that neither the BH mass, nor the accretion rate are defining factors in a QSO's radio-loudness.
    
    \item[-] The observed differences between the RL and RQ population are not gradual. Comparing composite spectra as a function of radio-loudness shows that all more radio-loud bins ($R_{2}$, $R_{3}$, ${R_{4}}$) differ from the radio-quietest ($R_{1}$), with features consistent with the classical RL and RQ division.

    \item[-] These changes cannot be explained by the addition of radio-undetected sources, as we have found that the radio-detected and radio-undetected quasars in $R_{1}$ are consistent with being drawn from the same parent distribution.
    
    \item[-] We have shown that RL quasars have on average higher SFRs than their RQ counterparts at any given redshift $0.5 < z < 1.9$, extending this result to significantly earlier cosmic epochs than previously known. The elevated levels of [O\,{\sc ii}] emission that we use to infer the SFRs have been corrected for possible influence of AGN contamination (following the procedure of \citealt{maddox2018ii}) and we have shown that our results cannot be explained by contamination from starburst galaxies.
    
\end{itemize}
Our results show that there is no clear-cut division in radio-loudness between RL and RQ quasars, and that the differences observed between them is not related to black hole mass or the accretion rate. As a result, we propose two distinct models: one requiring that RL quasars have rapidly spinning black holes in conjunction with abundant gas reservoirs, or are representatives of an earlier obscured phase of QSO evolution.

With the advent of future spectroscopic surveys such as WEAVE-LOFAR, the number of radio-loud QSOs will significantly increase. This will allow to us to investigate their spectral properties in greater detail and with higher significance. With higher S/N, however, we will need more sophisticated theoretical models to fit the composite spectra in order to disentangle their dust and host properties.

\section*{Acknowledgements}

We thank the anonymous referee for the helpful comments that have improved the manuscript.
MIA acknowledges support from the UK Science and Technology Facilities Council (STFC) studentship under the grant ST/V506709/1.
DJBS, MJH and ABD acknowledge support from the STFC under the grant ST/V000624/1. 
KJD acknowledges support from the STFC through an Ernest Rutherford Fellowship (grant number ST/W003120/1).
LKM is grateful for support from the Medical Research Council (MR/T042842/1). 
SD acknowledges support from the STFC via studentship grant number ST/W507490/1.
SS acknowledges support from the STFC via studentship grant number ST/X508408/1.
LOFAR is the Low Frequency Array designed and constructed by ASTRON. It has observing, data processing, and data storage facilities in several countries, which are owned by various parties (each with their own funding sources), and that are collectively operated by the ILT foundation under a joint scientific policy. The ILT resources have benefited from the following recent major funding sources: CNRS-INSU, Observatoire de Paris and Université d’Orléans, France; BMBF, MIWFNRW, MPG, Germany; Science Foundation Ireland (SFI), Department of Business, Enterprise and Innovation (DBEI), Ireland; NWO,
The Netherlands; The Science and Technology Facilities Council,
UK; Ministry of Science and Higher Education, Poland; The Istituto
Nazionale di Astrofisica (INAF), Italy.
This research made use of the Dutch national e-infrastructure with
support of the SURF Cooperative (e-infra 180169) and the LOFAR
e-infra group. The Jülich LOFAR Long Term Archive and the German LOFAR network are both coordinated and operated by the Jülich
Supercomputing Centre (JSC), and computing resources on the supercomputer JUWELS at JSC were provided by the Gauss Centre for
Supercomputing e.V. (grant CHTB00) through the John von Neumann Institute for Computing (NIC).
This research made use of the University of Hertfordshire high performance computing facility and the LOFAR-UK computing facility located at the University of Hertfordshire and supported by
STFC [ST/V002414/1], and of the Italian LOFAR IT computing infrastructure supported and operated by INAF, and by the Physics
Department of Turin University (under an agreement with Consorzio
Interuniversitario per la Fisica Spaziale) at the C3S Supercomputing
Centre, Italy.

\section*{Data Availability}
The data used in this work are publicly available and can be found at the LOFAR Surveys (\href{www.lofar-surveys.org}{www.lofar-surveys.org}) and the Sloan Digital Sky Survey website (\href{www.sdss.org}{www.sdss.org}.)



\bibliographystyle{mnras}
\bibliography{ref} 




\newpage
\onecolumn
\appendix

\section{Supplementary material}\label{sec:supplementary}

\begin{figure*}
\includegraphics[width=\textwidth]{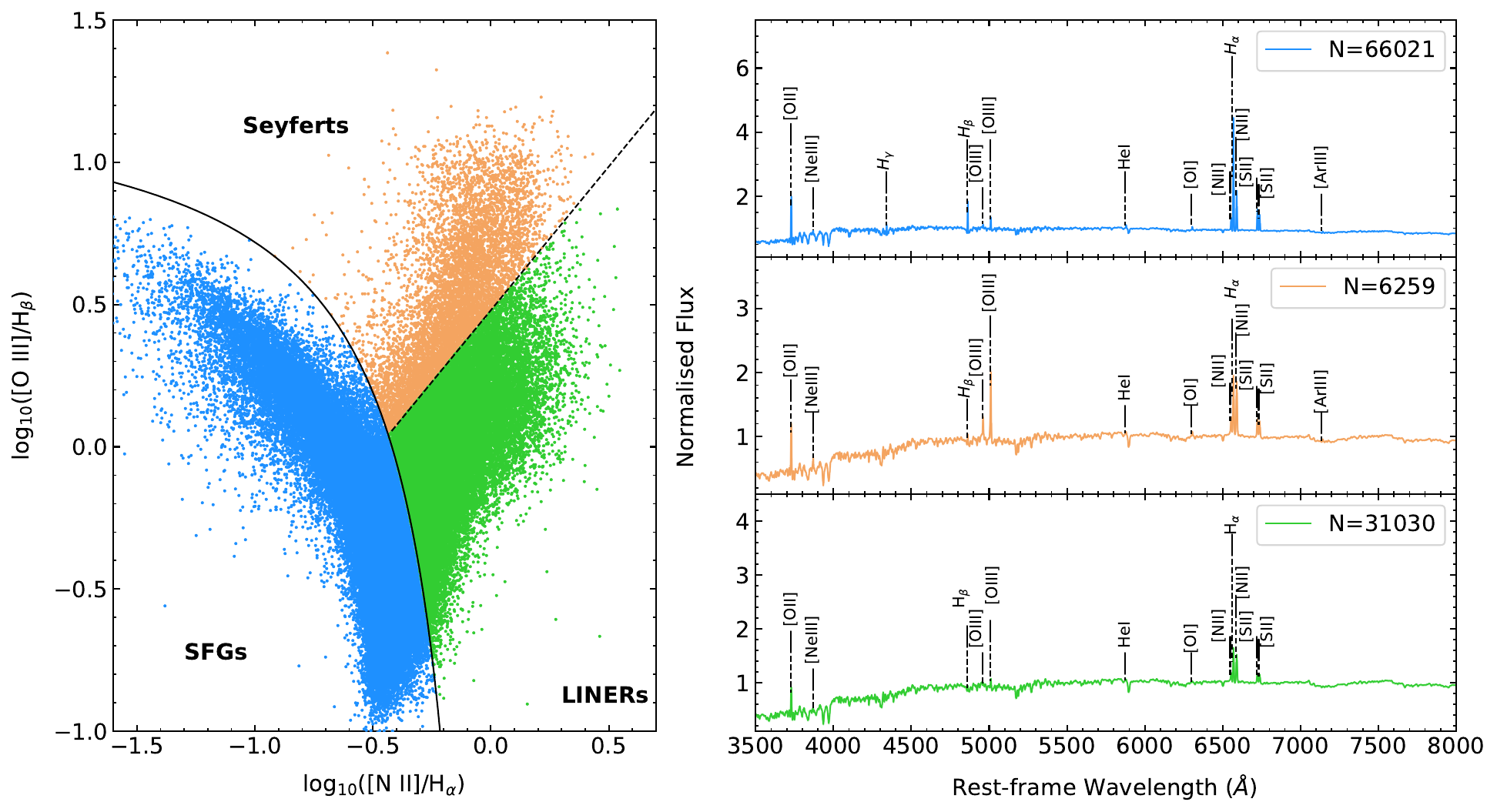}
    \caption{Results of applying our spectral stacking algorithm to sources according to their location on the BPT-NII diagram. \textbf{Left panel:} The BPT-NII diagram with dividing lines indicating  regions populated by SFGs (blue), Seyferts (orange) and LINERs (green) as defined relative to the \citet{kauffmann2003stellar} and \citet{kewley2006host} dividing lines (which are solid and dashed, respectively). \textbf{Right panel:} Composite spectra for each class colour-coded to match the regions of the BPT-NII diagram. Emission lines of interest are labelled, and the legend indicates the number of spectra included in each stack.}
    \label{fig:BPT stacks}
\end{figure*}

\begin{figure*}
\includegraphics[width=\textwidth]{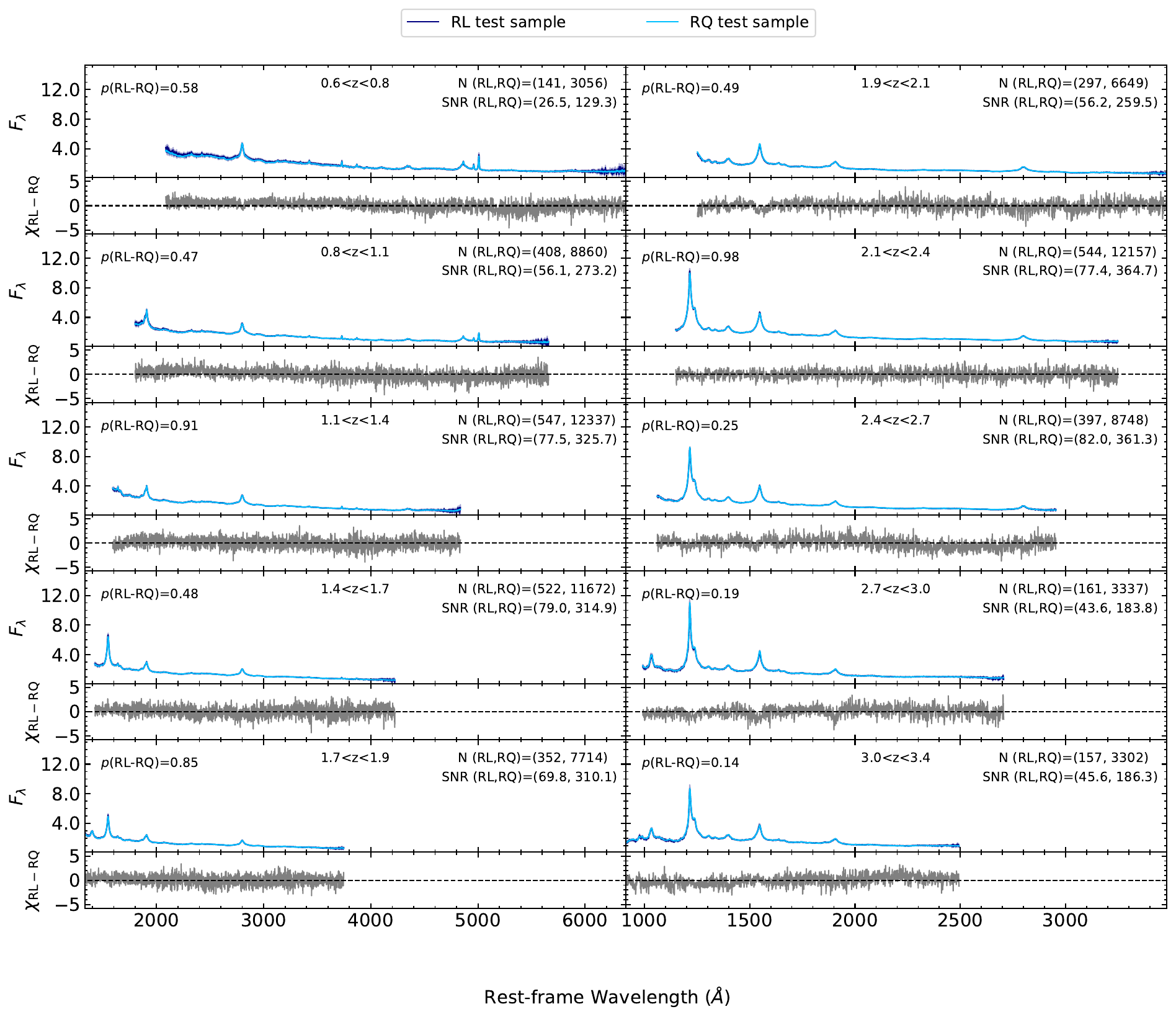}
    \caption{A Monte Carlo simulation representing the comparison between RL and RQ quasars under the null hypothesis that the RL population is drawn at random from the RQ population. The upper panel of each bin presents the composite spectra of the RL test sample (dark blue) and the RQ test sample (light blue), whereas the lower panel indicates the residual in units of propagated uncertainty (grey). The $p$-values for each of the null hypothesis tests are presented in the upper left corner, while the number of sources and median S/N are indicated in the top right corner of each panel.}
    \label{fig:RL_vs_RQ_NH}
\end{figure*}

\begin{figure*}
\includegraphics[width=\textwidth]{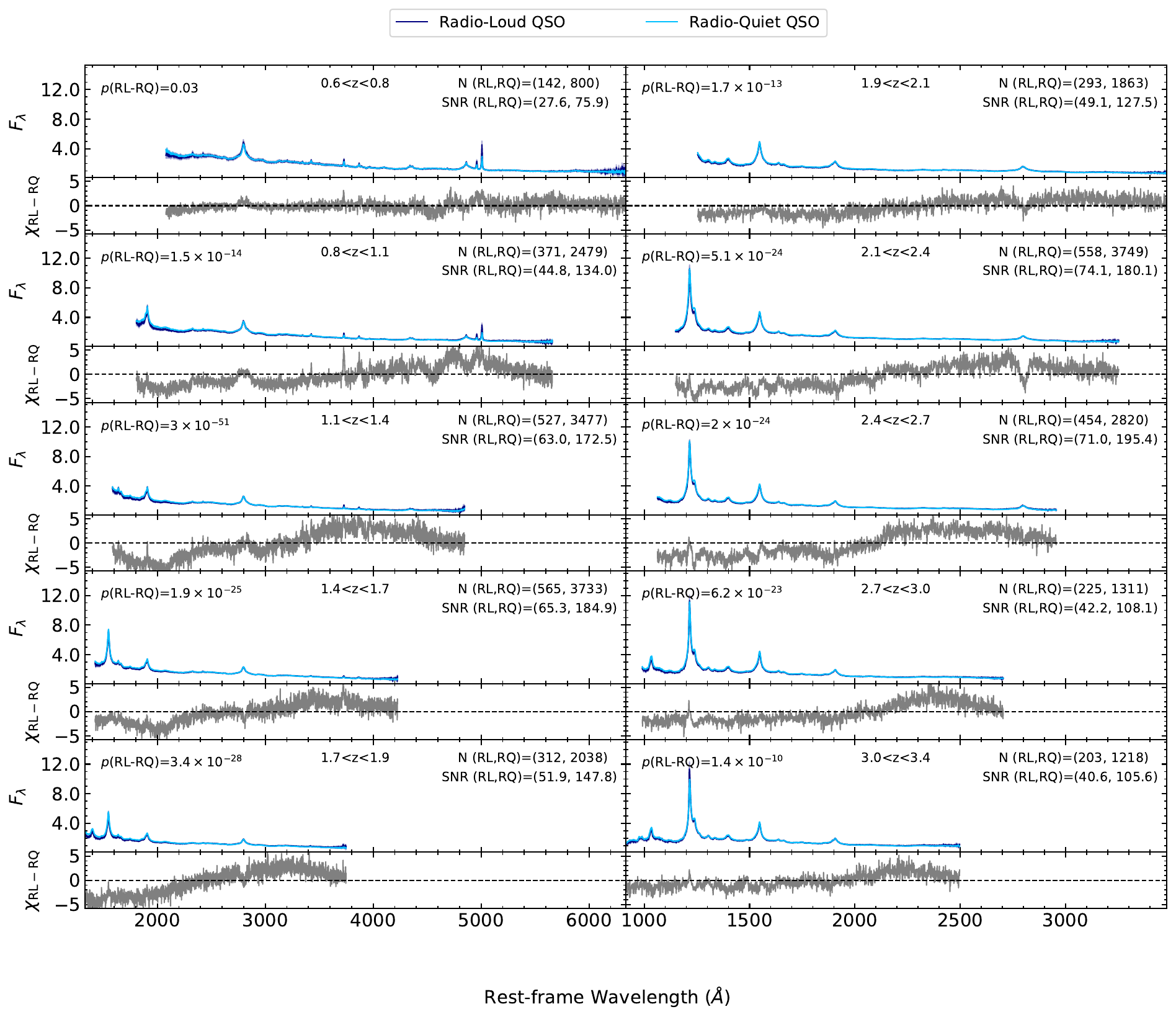}
    \caption{A high S/N comparison between the RL and RQ population matched in $z$, $M_{i}$ and $M_{\rm{BH}}$. As in Figure \ref{fig:RL_vs_RQ}, the upper panel of each bin presents the composite spectra of RL QSOs (dark blue) and RQ QSOs (light blue), whereas the lower panel indicates the residual in units of propagated uncertainty (grey). As for figure \ref{fig:RL_vs_RQ_NH}, $p$-values for each of the null hypothesis tests are presented in the upper left corner, while the number of sources and median S/N are indicated in the top right corner of each panel.}
    \label{fig:RL_vs_RQ_3d}
\end{figure*}

\begin{figure*}
\includegraphics[width=\textwidth]{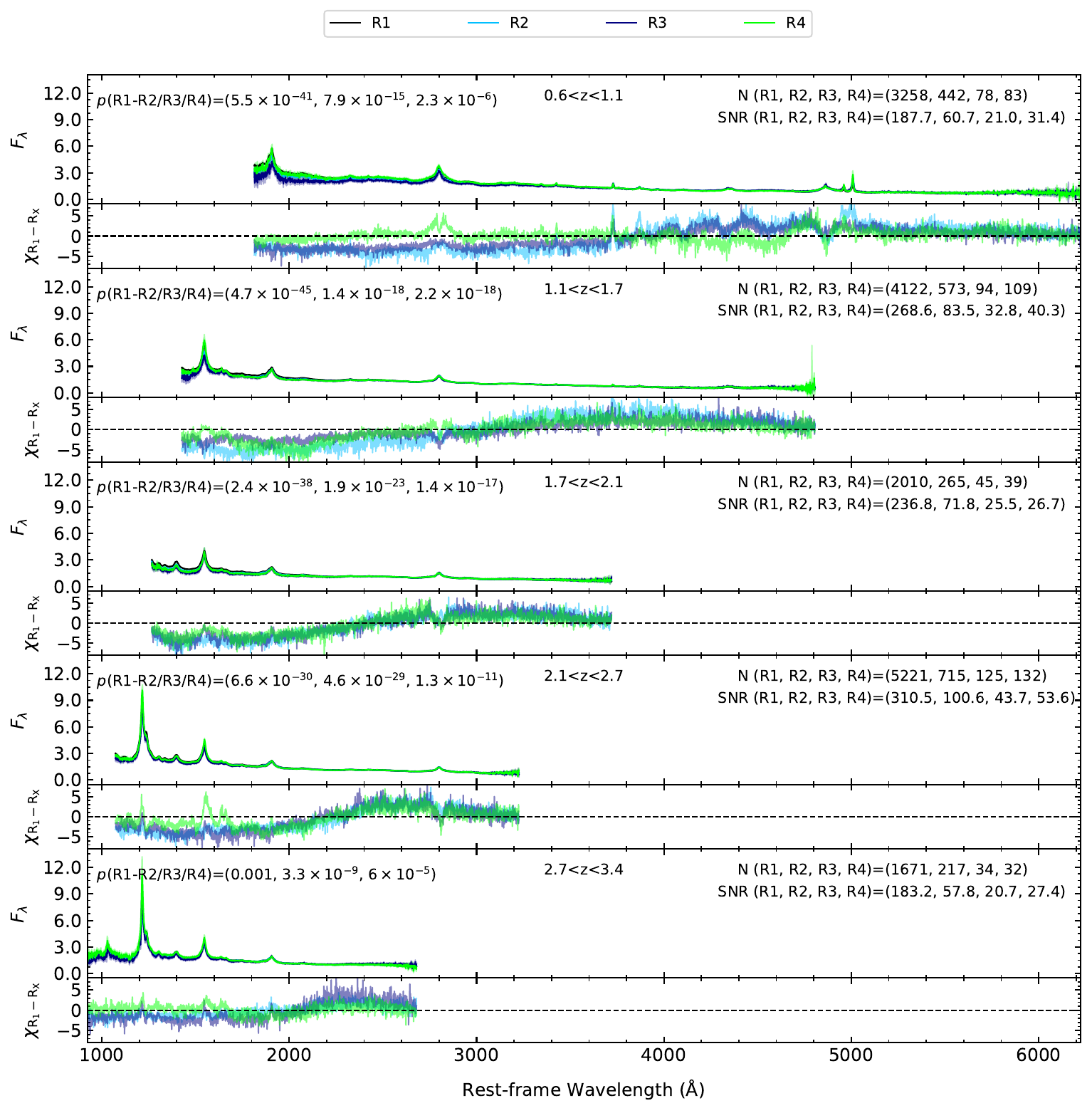}
    \caption{A high S/N comparison between the radio-quietest ($R_{1}$) and more radio-loud ($R_{2}$, $R_{3}$ and $R_{4}$) parts of the QSO population. The upper panel of each bin presents the composite spectra of $R_{1}$ (black), $R_{2}$ (light blue), $R_{3}$ (dark blue) and $R_{4}$ (green), whereas the lower panel indicates the residual in units of propagated uncertainty for each comparison. As before, the $p$-values for each of the null hypothesis tests are presented in the upper left corner, while the number of sources and median S/N are indicated in the top right corner of each panel.}
    \label{fig:Rbins}
\end{figure*}

\begin{figure*}
\includegraphics[width=\textwidth]{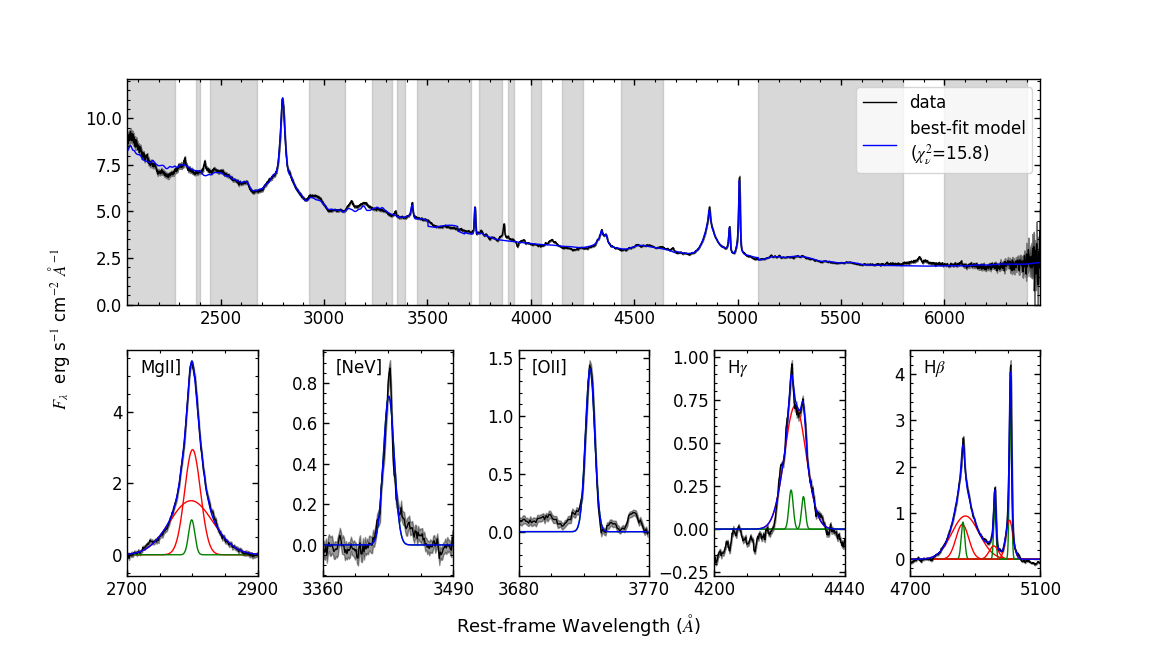}
    \caption{An example of the spectral fitting procedure with \textsc{\texttt{PyQSOFit}}. The top panel presents the composite spectra of the RQ population in the redshift range of $0.6<z<0.8$ (black) overlaid with the best fit model (blue). The grey shaded region show the wavelength windows used for the continuum fit. The lower panels show the best-fit model of individual line complexes (blue), along with the decomposition into broad (red) and narrow (green) line components.}
    \label{fig:RQ_fit}
\end{figure*}

\bsp	
\label{lastpage}
\end{document}